\documentclass[10pt,conference]{IEEEtran}

\usepackage{booktabs} 
\usepackage[flushleft]{threeparttable}
\usepackage{xcolor}
\usepackage[utf8]{inputenc}
\usepackage{float}
\usepackage{amssymb}
\usepackage{amsmath}
\usepackage{ifthen}
\usepackage{multirow} 
\usepackage{caption}
\usepackage{listings}
\usepackage[bookmarks=false]{hyperref}
\usepackage{url,moreverb,xspace}
\usepackage{tcolorbox}
\usepackage{xspace}
\usepackage{booktabs}

\usepackage{enumitem}
\usepackage{array,graphicx}
\usepackage{soul}
\usepackage{balance}
\usepackage{pifont}
\usepackage{etoolbox,siunitx}
\usepackage{nicefrac}
\usepackage{mathtools}

\usepackage{xcolor,pifont}
\newcommand*\colourcheck[1]{%
  \expandafter\newcommand\csname #1check\endcsname{\textcolor{#1}{\ding{52}}}%
}
\newcommand*\colourcross[1]{%
  \expandafter\newcommand\csname #1cross\endcsname{\textcolor{#1}{\ding{56}}}%
}
\usepackage[vlined, boxruled, linesnumbered] {algorithm2e}
\SetKw{KwBy}{by}
\SetKw{KwBreak}{break}
\SetKw{KwReturn}{return}
\def\HiLi{\leavevmode\rlap{\hbox to \hsize{\color{gray!35}\leaders\hrule height .8\baselineskip depth .5ex\hfill}}}

\usepackage{graphicx}
\usepackage{subcaption}
\usepackage{seqsplit}

\newcommand{\davetwo}{\mbox{DAVE-2}\xspace} %
\newcommand{\metric}{TSS\xspace} %
\newcommand{\onemetric}{OC-TSS\xspace} %

\newcommand{\changed}[1]{\textcolor{black}{#1}}
\newcommand{\changednew}[1]{\textcolor{black}{#1}}

\newcommand{\badcorrelation}[1]{\textcolor{red!70}{#1}}
\newcommand{\lowcorrelation}[1]{\textcolor{black!50}{#1}}

\SetKwComment{Comment}{$\triangleright$\ }{}
\SetCommentSty{itshape}
\newboolean{showcomments}
\setboolean{showcomments}{true}
\ifthenelse{\boolean{showcomments}}
{\newcommand{\nb}[2] {
  \fcolorbox{black}{gray!20}{\bfseries\sffamily\scriptsize#1:}
  {\sf\small$\blacktriangleright$\textit{#2}$\blacktriangleleft$}
}
}
{\newcommand{\nb}[2]{}
}

\newcommand{\head}[1]{\noindent\textbf{#1.}}

\newcounter{fcounter}
\newcommand{\curl}[1]{\footnote{\url{#1}}}

\makeatletter
\newcommand{\thickhline}{%
    \noalign {\ifnum 0=`}\fi \hrule height 1pt
    \futurelet \reserved@a \@xhline
}
\makeatother

\begin{document}
\pagenumbering{arabic} 
\pagestyle{plain}

\title{Assessing Quality Metrics for Neural Reality Gap Input Mitigation in Autonomous Driving Testing}

\author{\IEEEauthorblockN{Stefano Carlo Lambertenghi}
\IEEEauthorblockA{
\textit{Technical University of Munich, fortiss GmbH}\\
Munich, Germany \\
stefanocarlo.lambertenghi@tum.de, lambertenghi@fortiss.org}
\and
\IEEEauthorblockN{Andrea Stocco}
\IEEEauthorblockA{
\textit{Technical University of Munich, fortiss GmbH}\\
Munich, Germany \\
andrea.stocco@tum.de, stocco@fortiss.org}
}


\IEEEoverridecommandlockouts
\IEEEpubid{\makebox[\columnwidth]{XXX/\$31.00~\copyright2024 IEEE \hfill} \hspace{\columnsep}\makebox[\columnwidth]{ }}

\maketitle

\IEEEpubidadjcol

\begin{abstract}
Simulation-based testing of automated driving systems (ADS) is the industry standard, being a controlled, safe, and cost-effective alternative to real-world testing.
Despite these advantages, virtual simulations often fail to accurately replicate real-world conditions like image fidelity, texture representation, and environmental accuracy. This can lead to significant differences in ADS behavior between simulated and real-world domains, a phenomenon known as the sim2real gap.

Researchers have used Image-to-Image (I2I) neural translation to mitigate the sim2real gap, enhancing the realism of simulated environments by transforming synthetic data into more authentic representations of real-world conditions. However, while promising, these techniques may potentially introduce artifacts, distortions, or inconsistencies in the generated data that can affect the effectiveness of ADS testing. 

In our empirical study, we investigated how the quality of image-to-image (I2I) techniques influences the mitigation of the sim2real gap, using a set of established metrics from the literature. We evaluated two popular generative I2I architectures, pix2pix and CycleGAN, across two ADS perception tasks at a model level, namely vehicle detection and end-to-end lane keeping, using paired simulated and real-world datasets.
Our findings reveal that the effectiveness of I2I architectures varies across different ADS tasks, and existing evaluation metrics do not consistently align with the ADS behavior. 
Thus, we conducted task-specific fine-tuning of perception metrics, which yielded a stronger correlation. Our findings indicate that a perception metric that incorporates semantic elements, tailored to each task, can facilitate selecting the most appropriate I2I technique for a reliable assessment of the sim2real gap mitigation.
\end{abstract}

\begin{IEEEkeywords}
autonomous vehicles testing, generative adversarial networks, sim2real, reality gap.
\end{IEEEkeywords}

%
%




\section{Introduction}\label{sec:introduction}

Autonomous driving systems (ADS) are vehicles capable of traveling between destinations with little to no human intervention. 
Current ADS are equipped with advanced deep neural networks (DNNs) that process sensor data to perform different prediction and control tasks, such as lane keeping, object detection, object avoidance, and path planning~\cite{yurtsever2020survey,chen2023endtoend}. 

To validate the safety of ADS, the consolidated industrial practice by automakers includes collecting real-world driving data that is used within a multi-pillar approach that includes training driving policies, virtual testing, test track, and real-world testing~\cite{Cerf:2018:CSC:3181977.3177753}. 
Virtual testing uses simulation toolkits to assess the compliance of an ADS with the safety requirements on a wide range of virtual scenarios, some of which are extremely difficult to test in the real world~\cite{zhong2021survey,2020-Riccio-EMSE}. 

However, generalization to real-world conditions is not guaranteed~\cite{AfzalSimulation21}, as simulation platforms typically lack sensor fidelity and accurate representation of real-world conditions. This phenomenon, known as the reality gap or sim2real gap~\cite{chen2023endtoend,ADS_testing_survey_2022}, causes a mismatch between the behavior of autonomous vehicles in simulated environments and the real world. 
For ADS, understanding and mitigating the sim2real gap is crucial. When the ADS performance in the simulated environment does not directly translate to its performance on actual roads, it lowers the trustworthiness of testing and acceptance of ADS for large-scale deployment. 

In this work, we target the sim2real gap for vision-based ADS at a model level, i.e., considering pairs of simulated and real-world inputs to the ADS. 
In this context, data-driven neural solutions, specifically Image-to-Image translation models (referred to as I2I models hereafter), are used to transform input images pertaining to one domain (i.e., simulated driving scenes) into new images in a different domain (i.e., real-world driving scenes)~\cite{Kim2021_DriveGAN}. 

While I2I models have demonstrated remarkable performance in generating images that \textit{appear} realistic to human observers~\cite{DBLP:journals/corr/abs-1802-03446}, thus narrowing the inconsistent behavior between simulated and real-world data, they also bear notorious limitations in terms of the quality of their generated outputs, including issues like feature blending, color bleeding, or object omissions~\cite{chen2023endtoend}.
For this reason, it is important to ground the progress done in sim2real mitigation for ADS testing using established metrics that can measure the extent to which the gap between simulated and real-world data is reduced thanks to I2I models.

Unfortunately, there is still no consensus on which evaluation metrics are the most effective for this task. While studies~\cite{DBLP:journals/corr/abs-1802-03446,i2i-trans-metrics} have attempted to benchmark generative adversarial networks, including I2I models, additional research is required to determine whether these metrics are correlated with a reduction in sim2real gap of ADS that rely on neural-generated data as inputs. 

The objective of this paper is to systematically investigate the benefits and limitations of employing I2I models for mitigating the sim2real gap in the context of ADS testing. Our goal is to assess whether I2I models are a viable and effective means of bridging this gap. Our study has three primary objectives. 
First, we aim to assess whether the effectiveness of I2I models generalizes across various ADS tasks, or if certain I2I solutions are better suited for specific tasks. 
Second, we aim to assess the relationship between the existing metrics used to evaluate the quality of I2I model outputs and the performance, or lack thereof, of the ADS systems being tested. Finally, our third objective is to explore whether domain-specific fine-tuning of the existing metrics can improve their ability to correlate with the ADS behavior.

To address these objectives, we conducted a comprehensive evaluation involving two well-known I2I models, pix2pix and CycleGAN, in which we investigated their capabilities in mitigating the gap between simulation and reality for two critical ADS tasks, vehicle detection, and lane keeping. 
Additionally, we employed 13 existing metrics from the literature, both at the distribution level and at the single-image level, and assessed their correlation with the ADS behavior in terms of prediction error, confidence measured as a complement of DNN uncertainty, and attention error based on the heatmaps from the explainable artificial intelligence domain. Finally, we conducted task-specific fine-tuning of perception metrics, to explore whether this adaptation could improve its effectiveness in sim2real gap mitigation.

Our results revealed that the effectiveness of I2I models varies across different ADS tasks, and existing evaluation metrics do not consistently align with the ADS behavior. However, through task-specific fine-tuning of perception metrics, we achieved a more robust correlation than all other metrics of our study, for both I2I models and ADS tasks. This suggests that this task-specific perception metric can serve as a reliable indicator for assessing the reduction of the sim2real gap in model-level ADS testing.

Our paper makes the following contributions:

\begin{description}[noitemsep]
\item [Evaluation] An empirical validation of 13 existing metrics for I2I models in addressing the sim2real gap in ADS testing.
\item [Perception Metric] A task-specific perception metric to evaluate I2I models for sim2real scenarios in ADS testing.
\item [Dataset] An extension of an existing dataset for lane-keeping ADS simulator with precise segmentation maps.
\end{description}

\section{Image-to-Image Translation Models}\label{sec:i2i-models}

Generative models represent a class of machine learning models designed to produce novel data resembling a provided dataset. This is achieved by capturing the underlying patterns and structures within the data, enabling the generation of novel data samples.
Among these methods, Generative Adversarial Networks (GANs)~\cite{NIPS2014_5423} emerged as the most popular solution. GANs consist of two neural networks, namely the Generator and the Discriminator, that engage in a competitive learning process known as adversarial learning. The Generator's objective is to create synthetic data that is indistinguishable from real samples, while the Discriminator learns to accurately differentiate between them.

In the context of autonomous driving, GANs have been applied to various tasks such as input pre-processing, including raindrop removal~\cite{de_rain_1}, de-blurring~\cite{de_blur}, and de-hazing~\cite{de_haze}. They also found applications in sim2real I2I translation tasks~\cite{2023-Stocco-TSE,deeproad,sim2real_gans_cit,sim2real_gans_cit_2}, in which one significant challenge is ensuring that the generated images not only look realistic but also preserve the essential content and details of the input, i.e., their \textit{semantics}. Since adversarial training may not guarantee semantics preservation, researchers have proposed GAN variations that introduce constraints on the Generator's output. Instances of these solutions are pix2pix~\cite{pix2pix}, UNIT~\cite{unit}, DiscoGAN~\cite{disco_gan}, and CycleGAN~\cite{cyclegan}.

We focus our investigation on two popular I2I models, namely pix2pix~\cite{pix2pix} and CycleGAN~\cite{cyclegan}, for their significant impact (the original papers have more than 21$k$ citations each as of 16 January 2024), including software engineering applications for ADS testing~\cite{2023-Stocco-TSE,biagiola2023better,2023-Stocco-EMSE}. 

\head{pix2pix}
pix2pix is an I2I model based on conditional Generative Adversarial Networks (cGANs)~\cite{cgan}.
Unlike traditional GANs, the generator is provided with conditional information, such as an input image or label, in addition to random noise. 

In the case of pix2pix, the conditional information is an input image from the source domain (e.g., real-world driving images) that needs to be translated into the target domain (e.g., synthetic driving images), which forces the generator to produce data that aligns with the input. 
The discriminator, as in a GAN network, is tasked with distinguishing between a real target image and a generated one. However, in pix2pix, the adversarial loss drives the Generator towards producing realistic images, whereas the L1 loss (Mean Absolute Error) ensures pixel-level accuracy, enforcing the creation of contextually accurate generated images.

\head{CycleGAN}
CycleGAN targets unsupervised I2I translation, i.e., it does not require paired training data from the source and target domains. CycleGAN adopts four neural networks, two generators ($G_1$ and $G_2$) and two discriminators ($D_1$ and $D_2$), which are trained simultaneously.
The generator $G_1$ performs the sim2real translation from the source domain (e.g., simulated driving images) to the target domain (e.g., real-world driving images). Conversely, the generator $G_2$ performs the inverse real2sim transformation. The discriminator $D_1$ is trained to differentiate between real images from the target domain and images generated by $G_1$, whereas the discriminator $D_2$ fulfills a similar role for the inverse mapping.

While the generators $G_1$ and $G_2$ employ adversarial training to generate images that closely resemble images from the target domain, the preservation of the original image's content (e.g., vehicles, pedestrians, or traffic signs occurring in a real-life driving scenario) is not ensured. Thus, to ensure that the translation is both realistic and semantics-preserving, CycleGAN uses a cycle consistency loss to quantify the disparity between the source image and the image obtained by cycling through the translation process. In other words, when an image is translated from the source domain to the target domain and then back to the source domain, the resulting image should closely resemble the original source image.

\section{Image-to-Image Models Evaluation Metrics}\label{sec:metrics}

Researchers have identified issues within GANs, including I2I models, that could potentially affect systems reliant on the use of their generated images as input. These include visual anomalies such as color bleeding, feature blending, omission, hallucination, erroneous object textures, and pixel patterns which can introduce inaccuracies and inconsistencies, disrupting ADS related to object recognition, segmentation, and scene understanding tasks~\cite{gan_problems_1}. 

The objective of this study is to explore the advantages and drawbacks of employing I2I models for sim2real gap mitigation in the context of ADS testing. Our specific goals are twofold. First, we aim to assess whether the effectiveness of I2I models can be transferred to various ADS tasks or if certain I2I solutions are better suited for specific ones. Second, we aim to assess whether the existing metrics used to evaluate the quality of I2I model outputs are correlated with the performance of the ADS being tested and whether improvements to the existing metrics can be made. 

In this study, we evaluate the image quality assessment metrics provided by Borji~\cite{DBLP:journals/corr/abs-1802-03446} and by Pang et al.~\cite{i2i-trans-metrics}. 
The selected metrics pertain to two categories, namely distribution-level metrics and single-image metrics.

\subsection{Distribution-level Metrics}\label{sec:dl-metrics}

Distribution-level metrics are designed to evaluate the overall performance of an I2I translation model by considering the entire set of generated images as a collective output. These metrics provide insights into the global characteristics of the generated distribution, enabling them to work even if a mapping between the source and target domain is not available.

\head{Inception Score~\cite{inception-metric} (IS)}
This metric assesses the quality and diversity of image distributions as follows. IS employs the pre-trained Convolutional Neural Network (CNN) model, Inception~\cite{inception}, as a feature extractor to compute classification confidence (which is used as a proxy for quality) and entropy of the predicted class probabilities (which is considered as a measure of diversity) on all generated images. Therefore, IS consists of an aggregate value computed as the exponential of the entropy of the average class probabilities. 
A high IS score indicates high-quality and more diverse images.
    
\head{Fr{\'e}chet Inception Distance~\cite{fid-metric} (FID)}
This metric quantifies the distance between the distributions of features extracted from real and generated images.
Similarly to IS, the FID metric relies on a pre-trained Inception model as a feature extractor. 
The extracted features of both real and generated images are then modeled as a multivariate Gaussian Distribution and compared using the Fr{\'e}chet distance~\cite{frechet}.
A lower distance should indicate high similarity between the real and generated distributions, indicating a well-performing generative model.
    
\head{Kernel Inception Distance~\cite{kid-citations} (KID)}
This metric quantifies the similarity between the distributions of feature representations by modeling them as kernel mean embeddings. The Gram matrix~\cite{gram_matrix} is computed for both the real and generated images to capture relationships between their features. The square root of Maximum Mean Discrepancy (MMD)~\cite{maximum-mean} is then used to quantify the difference between the distributions based on the Gram matrices. A lower MMD score should indicate a better alignment between the two image distributions.

\subsection{Single-Image Metrics}

Single-image metrics focus on evaluating the quality of individual images generated by a model. These metrics assess how well each output image aligns with a ground truth or reference image.
Due to the direct comparison between two images, these techniques require the input and target domain to have an explicit mapping.
We categorize single-image metrics in \textit{content-based metrics} and \textit{perception-based metrics}. 

\subsubsection{Content-based Single-Image Metrics}

These metrics only use the input image pair attributes to evaluate image differences. We consider seven content-based metrics.

\head{Structural Similarity Index (SSIM)~\cite{struct_sim}} 
It evaluates the quality of a generated image by comparing its structural information, such as edges, textures, and patterns, to a reference image.
The SSIM output is obtained by a combination of luminance (the difference in average intensity values), contrast (the distance between the image contrast distributions), and structure (the normalized covariance between the pixel intensities of the two images).
A higher SSIM value indicates greater similarity between the two images in terms of their structural information, suggesting higher image quality.

\head{Peak signal-to-noise ratio (PSNR)~\cite{inception-metric}} 
It measures the difference between an image pair by calculating the ratio of the maximum possible pixel value (peak signal) to the Mean Square Error (MSE) between the reference image and the reconstructed/generated image pixel values (noise). 
A high PSNR value should indicate a low average error between the two image domains thanks to a well-performing model.

\head{Mean Squared Error (MSE)} 
It quantifies the average squared difference between the pixel values of two images. Lower MSE values indicate higher similarity at the pixel level between the generated and the reference image~\cite{struct_sim}.

\head{Cosine Similarity (CS)~\cite{cosine_sim}}
It is a metric used to quantify the similarity between two feature representations. It calculates the cosine similarity of the feature vectors extracted using the same pre-trained classification models as in CPL. A higher cosine similarity value suggests that the images are more similar in terms of the features.

\head{Texture Similarity Index (TSI)} 
It is used to measure the texture properties of a generated and real image pair.
For each of the two inputs, multiple Gray-Level Co-occurrence Matrices (GLCMs)~\cite{glcm} are computed and used to capture spatial relationships between pixel values.
Using the output of these matrices it is possible to analyze how the textures differ in terms of properties like contrast, dissimilarity, homogeneity, energy, and correlation.
The metric output is an aggregate of the absolute difference in texture properties between the two images. A lower texture similarity score suggests two images with close texture properties.

\head{Wasserstein Score (WD)~\cite{wd_loss}}
Also known as Earth Mover's Distance (EMD)~\cite{earthmover}, it assesses the similarity between two images by considering the cost of transforming one image into the other. It quantifies the minimum amount of work (or distance) required to redistribute the mass from the pixels in the generated image to match the reference image. A lower Wasserstein Score indicates greater similarity.

\head{KL Divergence (KL)~\cite{kl_div}}
It is a measure of the difference between two probability distributions. In the context of images, it quantifies the difference in pixel value distribution between a generated image and a reference image. A lower KL Divergence indicates greater similarity in pixel value distribution.

\head{Histogram Intersection (HistI)~\cite{hist_inter}} 
It measures the similarity between two images by computing the intersection of their histograms, i.e., the distribution of pixel values. A higher intersection value suggests greater similarity.

\subsubsection{Perception-based Single-Image Metrics}

These metrics evaluate the differences between a real and a generated image by observing high-level features computed leveraging a pre-trained model for feature extraction, such as ResNet50~\cite{resnet} or VGG~\cite{VGGNet}.
We consider three perception-based metrics.
\head{\\Classifier Perceptual Loss (CPL)~\cite{perc_loss}} 
It measures the difference between the feature maps created by the convolutional layers of a pre-trained classification model. 
Feature maps capture high-level representations of the input images, such as textures, objects, and shapes. By comparing the feature maps of a generated image to those of a reference image, perceptual loss aims to measure if the generated image not only matches pixel values but also retains analogous perceptual features. A lower perceptual loss value indicates that two images share more perceptual characteristics, suggesting high similarity.

\head{Semantic Segmentation Score (SSS)} 
It measures the semantic details and the structural difference between two images by applying a pre-trained segmentation architecture such as FCN-8~\cite{fcn_8} on the generated image, and by comparing its output with the ground truth segmentation label of the source image. A high error between the semantic representation of the original and the predicted image should indicate a lack of fidelity in terms of image content.

\section{Empirical Study}\label{sec:empirical-study}

The goal of our empirical study is to evaluate different I2I models for sim2real gap mitigation in the context of ADS testing. We consider two popular, yet different tasks related to autonomous driving, namely vehicle detection and lane keeping.
These components are deemed essential by the U.S. Department of Transportation, National Highway Traffic Safety Administration (NHTSA), but they require rigorous testing to ensure that the vehicle can effectively stay in its lane and recognize pertinent objects before deployment~\cite{precrashreport}.

\subsection{Research Questions}\label{sec:rqs}

We consider the following research questions.

\noindent
\textbf{RQ\textsubscript{1} (Transferability):} \textit{Does the effectiveness of I2I models for sim2real mitigation vary for different tasks?}

The first research question evaluates the effectiveness of I2I models for sim2real gap mitigation in the context of the two considered ADS testing tasks.

\noindent
\textbf{RQ\textsubscript{2} (Correlation):} \textit{How do existing I2I evaluation metrics correlate with the associated ADS behavior?}

In the second research question, we aim to assess whether our evaluated metrics are informative with respect to the (mis)behavior of ADS that use these inputs for sim2real gap mitigation. 

\noindent
\textbf{RQ\textsubscript{3} (Fine-tuning):} \textit{Does fine-tuning of I2I perception-based metrics improve the sim2real mitigation task?}

Finally, in the third research question, we further fine-tune the perception metrics on each individual task to assess whether improvements can be achieved beyond the original metric propositions.

\begin{figure*}
  \centering

  \begin{subfigure}{0.49\textwidth}
    \begin{subfigure}{0.49\linewidth}
      \centering
      \includegraphics[width=0.97\linewidth]{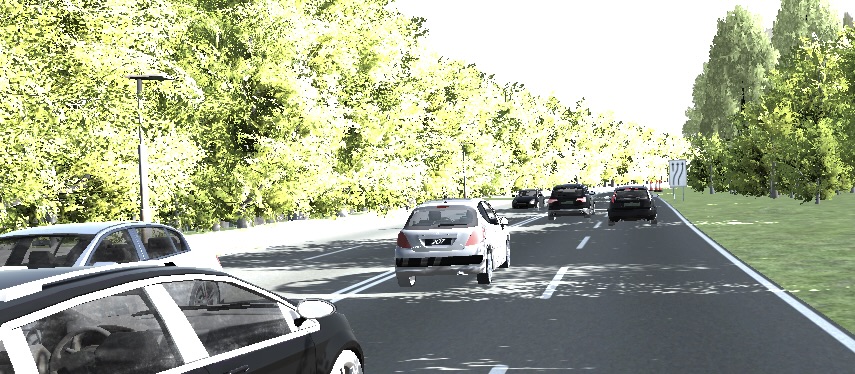}
      \caption{Vehicle Detection Sim.}
      \label{fig:subfig1}
    \end{subfigure}
    \begin{subfigure}{0.49\linewidth}
      \centering
      \includegraphics[width=0.97\linewidth]{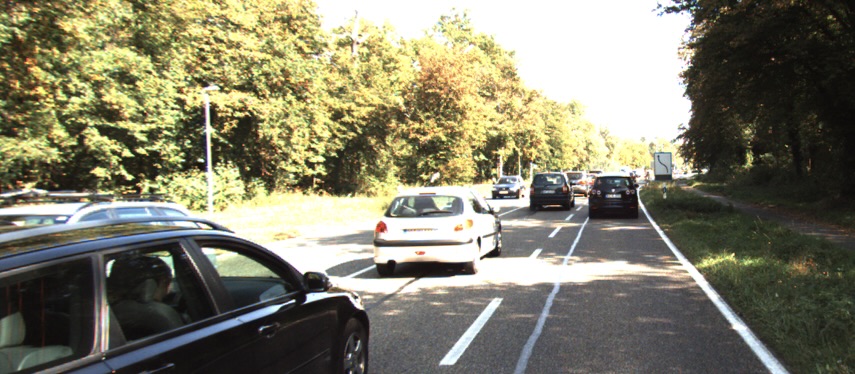}
      \caption{Vehicle Detection Real.}
      \label{fig:subfig2}
    \end{subfigure}
  \end{subfigure}
  \begin{subfigure}{0.49\textwidth}
    \begin{subfigure}{0.49\linewidth}
      \centering
      \includegraphics[width=0.97\linewidth]{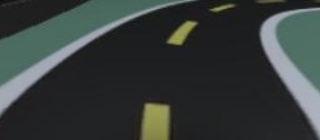}
      \caption{Lane Keeping Sim.}
      \label{fig:subfig3}
    \end{subfigure}%
    \begin{subfigure}{0.49\linewidth}
      \centering
      \includegraphics[width=0.97\linewidth]{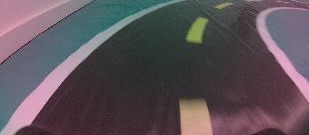}
      \caption{Lane Keeping Real.}
      \label{fig:subfig4}
    \end{subfigure}
  \end{subfigure}

  \caption{An example of dataset sim2real pairs for the Vehicle Detection (a-b) and Lane Keeping (c-d) datasets.}
  \label{fig:dataset_paired}
\end{figure*}

\subsection{Datasets}

\subsubsection{Vehicle Detection Dataset}

To perform translation quality evaluations for sim2real vehicle detection tasks, we adopt the KITTI~\cite{kitti} and vKITTI~\cite{vkitti} datasets. We use vKITTI (sim) and KITTI (real) as source and target datasets for I2I models (\autoref{fig:dataset_paired}(a-b) shows an example).

More in detail, the KITTI dataset (Karlsruhe Institute of Technology and Toyota Technological Institute) is a widely used benchmark for autonomous driving perception research~\cite{kitti_use_1,kitti_use_2,yolo}. It provides a large collection of real-world data (e.g., GPS, image, and LIDAR sensors) captured from a moving vehicle in multiple road environments. The ground truth is represented by bounding boxes (i.e., rectangular regions) that specify the position and extent of the objects of interest. Additionally, semantic segmentation maps assign specific object class labels to individual pixels, allowing for pixel-level identification of different object categories (e.g., vehicles, road pavement).
The vKITTI dataset (Virtual KITTI) is a virtual replica of a subset of KITTI, rendered using the Unity computer graphics engine~\cite{unity}. The dataset was obtained by performing a semi-supervised real-to-virtual world cloning method where object positions and orientations in the virtual world were calculated based on their relationships with the camera and road positions in KITTI. Differently, environmental objects such as trees, buildings, and secondary roads were manually placed in the environment~\cite{vkitti}.
The final dataset includes 1,064 sim2real image pairs and labels across KITTI and vKITTI.

\subsubsection{Lane Keeping Dataset}

The second dataset involves imitation-learning-based end-to-end lane-keeping across virtual and real-world environments. We have considered the datasets from recent work on sim2real gap estimation for ADS driving~\cite{2023-Stocco-TSE}, in which the authors provide a dataset of images captured with a physical small-scale autonomous vehicle performing lane-keeping on a real-world indoor track and the associated driving simulator (\autoref{fig:dataset_paired}(c-d) shows an example).
Specifically, the data is collected using the DonkeyCar~\cite{donkeycar}, a framework that includes a 1/16th scale vehicle and a digital twin simulation platform (\autoref{fig:dataset_paired}(c-d) shows an example). In total, the dataset includes 14$k$ simulated driving frames, of which 5$k$ are unlabeled and 9$k$ labeled, and 21$k$ driving frames, of which 6$k$ unlabeled and 15$k$ labeled.

\subsection{ADS Under Test}

\subsubsection{Vehicle Detection}

Concerning the vehicle detection task, we adopted the YOLOv3~\cite{yolov3} model (You Only Look Once v.3), an object detection and classification CNN architecture that has been adopted in the autonomous driving domain due to its accuracy and performance on constrained hardware~\cite{yolo_use_1,yolo_use_2}.
We chose version 3 for its maturity, its availability of robust pre-trained model weights~\cite{yolo_weights_web_1,yolo_weights_web_2}, and its higher inference speed compared to more recent variations~\cite{yolo_v3_speed}.
In this study, we use a model pre-trained on the Microsoft Common Objects in Context dataset (COCO)~\cite{coco}, an image dataset with semantic labels for 50 object classes and 123k samples, of which 13k pertain to vehicles.
The model outputs for each detected object are the object class, the confidence of said classification, and the bounding box containing the object.
For this study, we only consider object predictions pertaining to vehicles due to a marginal presence of pedestrians, bicycles, and other classes relevant to ADS tasks in the subset of (v)KITTI. In the rest of the paper, we shall use the terms object/vehicle detection dataset interchangeably. 
\subsubsection{Lane Keeping}

\davetwo is a CNN model for lane-keeping tasks based on imitation learning~\cite{nvidia-dave2}.
Given an input image representing a road scene, \davetwo processes it through three convolutional layers for feature extraction followed by five fully connected layers for steering angle regression. By providing recorded input images associated with steering commands, this model is capable of learning how to keep a vehicle on the road.
Due to its performance despite its simple architecture, this model has been used for a large variety of ADS testing studies~\cite{2023-Stocco-TSE,biagiola2023better,deeptest}. More importantly, \davetwo exhibits realistic behaviors in simulated environments, with performance degradation~\cite{2020-Humbatova-ICSE} that closely mimics reality when the generated images exhibit distortions in the lanes~\cite{nvidia_cnn_vision}. 

\subsection{ADS behavior Metrics}\label{ads_bvh_metrics}

To evaluate the behavior of both ADS, we conduct a comparison between ground truth annotations of real-world images and the ADS outputs generated using simulated, I2I translated, and corresponding real images. 
This evaluation is based on three ADS behavioral metrics~\cite{2021-Jahangirova-ICST}: prediction error, confidence, and attention error. 

\head{Prediction Error}
The first behavioral metric is prediction error, i.e., the discrepancy between the predicted output of the ADS and the ground truth value. 
Intuitively, when the I2I model generates poor translations, it leads to high ADS prediction errors, signifying ineffective sim2real gap mitigation, while high-quality translations result in low prediction errors, indicating successful mitigation.

For YOLOv3, we use the Intersection over Union (IoU)~\cite{iou} metric to assess the alignment between predicted bounding boxes and the ground truth bounding boxes.
Any significant misalignment leads to false positives and false negatives, whose severity is further assessed by considering their respective areas. 
For example, a false negative that involves a vehicle in the foreground is deemed more critical than one in the background. For instances in which the bounding boxes are correctly matched, i.e., a successful object detection, we also measure the area difference between the two boxes to quantify the degree of bounding box misalignment. This methodology enables us to assess the accuracy of vehicle detection as well as to evaluate the extent of the bounding boxes' error prediction. 
For \davetwo, we compute the MSE between the predicted and ground truth steering angle.

\head{Confidence}
The second behavioral metric consists of a measure of confidence associated with the ADS prediction. 
We follow the intuition that this metric can help evaluate the quality of I2I translated inputs as high confidence may suggest a high degree of faith in the accuracy of the transformation, whereas low confidence may indicate a higher likelihood of discrepancies between real and translated images. 
In YOLOv3, the confidence of class predictions is readily available as one of the outputs. To assess the overall confidence of the model for one input, we simply aggregate the confidence scores for all predicted vehicles, resulting in a single averaged confidence value for the vehicle class. 
In contrast, the standard configuration of DAVE-2 provides only inference outputs, i.e., steering angles. Thus, following existing guidelines~\cite{Weiss2021FailSafe,2024-Grewal-ICST}, we use the predictive variance of dropout-based DNNs called Monte Carlo dropout, or MC-Dropout (MCD)~\cite{10.5555/3045390.3045502}. MCD leverages dropout layers, where random neurons are deactivated during inference, to inject randomness into the ADS predictions. By carrying out multiple iterations with dropout applied, MCD generates a range of predictions for the same input. The variance of the observed probability distribution quantifies the uncertainty (i.e., the confidence error).

\head{Attention Error}
The last ADS behavioral metric relies on attention maps generated through explainable artificial intelligence techniques (XAI). Attention maps are post-training methods that highlight the specific input pixels that exert the most significant influence on the output predictions~\cite{2022-Stocco-ASE}.
In our study, we employ the GradCam algorithm~\cite{DBLP:journals/corr/SelvarajuDVCPB16}, a widely used technique based on gradient back-propagation which enables us to gain insights into the image areas that the ADS considers relevant during its execution.
We utilize the attention map as a quality indicator to determine if the ADS is directing its attention to image regions pertinent to its intended task. When the attention map indicates ADS focus on areas different from those obtained on real-world data, it could indicate that the translation process has impacted the ADS reliability with respect to the unaltered real-world data.

For each input image, we use as ground truth the corresponding real-world semantic segmentation map, which provides a pixel-level identification of different objects and regions within the image. We use it to automatically detect the areas of interest (i.e., the vehicles for vehicle detection and road lanes for lane keeping) and compare them with the scores of the attention map generated by the I2I model for that image. Specifically, we apply a binarization process to both maps. For attention maps, we employ a threshold that highlights image areas of significant importance to the model, retaining those with an intensity above 50\%. 
In the semantic segmentation map, we assign a value of 1 to pixels that represent relevant objects, while setting all other pixels to 0. Subsequently, we compute the MSE between these two binary maps to quantify the model attention error.

\begin{table}[t]
\caption{Datasets Usage (\textcolor{green}{\ding{52}} used; \textcolor{red}{\ding{56}} not used).}\label{tab:table_datasets}

\footnotesize

\setlength{\tabcolsep}{1.4pt}
\renewcommand{\arraystretch}{1.1}

\begin{tabular}{@{}llllccc@{}}

\toprule

&\bf  Data split (set) & \begin{tabular}[c]{@{}c@{}}\bf Used\\ \bf for\end{tabular} & \begin{tabular}[c]{@{}c@{}}\bf Size\\  {[}\bf sim/real{]}\end{tabular} & \begin{tabular}[c]{@{}c@{}}\bf ADS\\ \bf labels\end{tabular} & \bf Paired & \begin{tabular}[c]{@{}c@{}}\bf Seg.\\ \bf maps\end{tabular} \\ 
 
\midrule

\multirow{2}{*}{Vehicle Detection} & Implementation & I2I/Seg & {[}1,064/1,064{]}  & \textcolor{green}{\ding{52}}   & \textcolor{green}{\ding{52}}    & \textcolor{green}{\ding{52}}                                                         \\ 
                              & Evaluation     & RQs & {[}1,064/1,064{]} & \textcolor{green}{\ding{52}} & \textcolor{green}{\ding{52}} & \textcolor{green}{\ding{52}} \\ [0.5em]

\multirow{3}{*}{Lane Keeping} & Train          & ADS & {[}-/7905{]} & \textcolor{green}{\ding{52}} & \textcolor{red}{\ding{56}}  & \textcolor{red}{\ding{56}}  \\
                              & Implementation & I2I/Seg & {[}5,361/5,361{]} & \textcolor{red}{\ding{56}} & \textcolor{red}{\ding{56}}  & \textcolor{green}{\ding{52}} \\
                              & Evaluation     & RQs & {[}197/197{]}  & \textcolor{green}{\ding{52}} & \textcolor{green}{\ding{52}} & \textcolor{green}{\ding{52}} \\ 
                              
\bottomrule

\end{tabular}
\end{table}

\subsection{Procedure}
\head{Dataset Pairing, Pre-processing, and Splitting}
Our study requires splitting the simulated and real datasets into different independent sets, with different requirements in terms of pairing. \autoref{tab:table_datasets} summarizes the dataset size and split used in the paper. 
For evaluating the I2I models in relation to the ADS behavior metrics, our study requires an \textit{evaluation set} of \textit{paired} simulated and real data. Without this consistency, assessing whether the model successfully maintains content while improving quality and style becomes challenging. 
A separate \textit{implementation set}, for which no pairing is required, is used to train the I2I models as well as the segmentation model used in RQ\textsubscript{3}. Finally, the lane-keeping case study also requires a \textit{training set} for training \davetwo.

The vehicle detection dataset already meets the pairing requirement as vKITTI is a synthetic counterpart of a subset from the KITTI dataset. Thus, we divided the paired data into two $\approx$1,000-image sets for the \textit{implementation set} and \textit{evaluation set}, without further preprocessing.
In contrast, for the lane-keeping dataset, we established a paired \textit{evaluation set} of simulated and real labeled images from the original test set~\cite{2023-Stocco-TSE} by performing a manual alignment of 197 frames between simulated and real images.\footnote{While the size of the evaluation set of the lane-keeping dataset is lower than the one for the vehicle detection dataset, the number of images is sufficient to represent all the conditions of the real-world road track used in the previous paper~\cite{2023-Stocco-TSE}, an 11 m long track, 52 cm wide. Clockwise, the track features three bends on the right and one on the left.}
The remaining (unpaired) data was divided into two groups: a \textit{training set} of 7,905 labeled real images and an \textit{implementation set} of 5,361 unlabeled simulated and real images.

To train pix2pix, due to its supervised nature, we need a substantial number of paired images. 
The vehicle detection dataset already includes the required semantic maps for pix2pix. In contrast, for the lane-keeping dataset, we generated segmentation maps by associating specific pixel values with semantic classes in each image (i.e., road and background), followed by manual refinement to ensure map accuracy. 

\begin{figure*}[t]
  \centering
  \begin{subfigure}{\textwidth}
  \includegraphics[width=\linewidth]{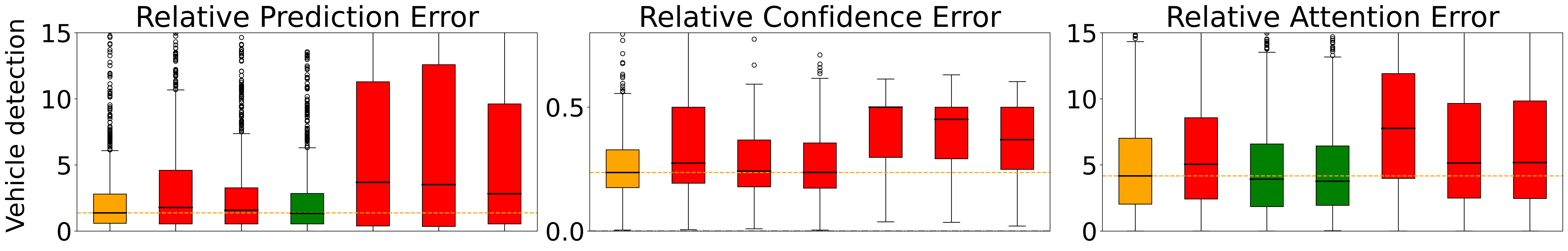}\\
  \includegraphics[width=\linewidth]{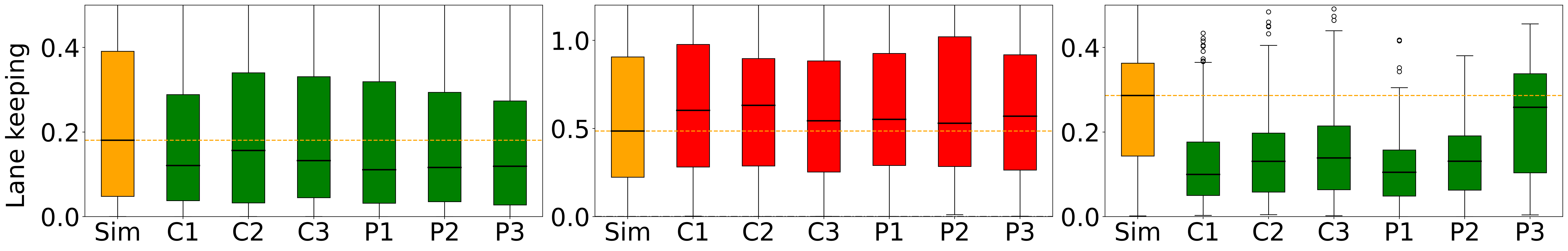}
  \end{subfigure}
  \caption{Distributions of ADS behavior on different image domains. \changed{The black horizontal line represents the \textit{median} of the distribution, whereas the color indicates whether the median is lower (green) or same/higher (red) as compared to the simulated domain (yellow); In each boxplot, distributions with lower values are preferable. The figure is best viewed in color.}}
  \label{fig:rq1}
\end{figure*}

\head{Training I2I Models and Data Generation\label{train_i2i}}
To train both pix2pix and CycleGAN, we used the \textit{implementation set}, as follows. pix2pix is trained using real images and their corresponding segmentation masks, whereas CycleGAN is trained using unpaired simulated and real images. We trained pix2pix for $\approx$23 hours and CycleGAN for $\approx$2.5 hours on a Google Cloud TPU v4 with 36GB of VRAM equivalent~\cite{google_tpu}.
\changed{Both models are trained using the original code provided by the authors.} 
During training, we saved the checkpoints (i.e., model weights) at the end of each epoch, which allowed us to consider models at various stages of maturity. After training, we selected three checkpoints for pix2pix and CycleGAN, approximately at 10-50-90\% of the training time, to assess poorly/medium/largely trained models (\autoref{tab:table_model_epochs_all}). Their respective generators were then used to produce a translated copy of the simulated data in the \textit{evaluation set}.

\begin{table}[h]
\caption{Training statistics for the I2I models.}\label{tab:table_model_epochs_all}
\footnotesize
\setlength{\tabcolsep}{1.4pt}
\renewcommand{\arraystretch}{1.1}

\begin{tabular}{llllllll}
\toprule
&  \multicolumn{3}{c}{\bf Training} && \multicolumn{3}{c}{\bf Checkpoints} \\
\cmidrule{2-4}
\cmidrule{6-8}
 
& \begin{tabular}[c]{@{}l@{}}I2I model\end{tabular} & \begin{tabular}[c]{@{}l@{}}Epochs\end{tabular} & \begin{tabular}[c]{@{}l@{}}Time (h)\end{tabular} & & \begin{tabular}[c]{@{}l@{}}Ckp. name\end{tabular} & Epochs & 

\begin{tabular}[c]{@{}l@{}}Time (h)\end{tabular} \\

\midrule
 
\multirow{2}{*}{\begin{tabular}[c]{@{}l@{}}Vehicle\\ Detection\end{tabular}} & pix2pix & 130 & 23.1 & & P1/P2/P3 & 14/70/120 & 2.4/12.4/21.3 \\
 & CycleGAN &  38 & 2.63 & & C1/C2/C3 & 4/19/33 & 0.2/1.3/2.2 \\ [0.5em]
 
\multirow{2}{*}{\begin{tabular}[c]{@{}l@{}}Lane\\ Keeping\end{tabular}} & pix2pix  & 170 & 22.95 & & P1/P2/P3 & 20/80/150 & 2.7/10.8/20.2 \\
 & CycleGAN  & 40 & 2.42 & & C1/C2/C3 & 8/20/34 & 0.4/1.2/2 \\
\bottomrule
\end{tabular}%
\end{table}

\head{Training ADS}
The upper bound number of epochs was set to 200, with a batch size of 128 images and a learning rate of 0.001. We used early stopping with a patience of 10 epochs on the validation set. The DNNs use the Adam optimizer~\cite{kingma2014adam} to minimize the MSE between the predicted steering angles and the ground truth value. 

\head{RQ\textsubscript{1} (Transferability)}
To answer RQ\textsubscript{1}, for each ADS task, we execute the ADS models (YOLOv3 and \davetwo) on the real-world \textit{evaluation set} to collect the nominal prediction error, confidence values, and attention errors (\autoref{ads_bvh_metrics}). 
We then executed each ADS on the simulated \textit{implementation set} and on the corresponding data generated by the six I2I translation models created for each task (three for pix2pix and three for CycleGAN) and collected the corresponding ADS behavior metrics. Finally, we computed the relative error concerning the score of these metrics on real-world data, which provided a quantification of how the ADS behavior deviated from the expected behavior. 

\head{RQ\textsubscript{2} (Correlation)}
To address RQ\textsubscript{2}, for both ADS tasks, we compute the metrics outlined in \autoref{sec:dl-metrics} on the I2I translated images from the evaluation set used for RQ\textsubscript{1}.
Then, we compute the correlation between \changed{these} metrics and the relative errors of ADS behaviors using Pearson's correlation coefficient~\cite{pearson}. 
For distribution-level metrics, we compute the correlation with the average of the \changed{relative} ADS behavioral metrics. For single-image metrics, we compute their correlation with the \changed{relative} ADS behavior for each input domain separately. To compute IS and FID, we utilize InceptionV3~\cite{inception_v3}, whereas for computing CPL, CS, and SSS, we used a VGG16~\cite{VGGNet} model, both pre-trained on ImageNet~\cite{imagenet}.

\head{RQ\textsubscript{3} (Fine-tuning)}
We develop a perception metric customized for our image domains through fine-tuning. Particularly, we leverage a SegFormer model~\cite{SegFormer} pre-trained on the Cityscapes dataset~\cite{cityscapes}, a transformer-based segmentation model, renowned for image semantic segmentation tasks.
\changed{Which is fine-tuned to the specific ADS tasks by training it only using} segmentation masks that contain the classes relevant to each ADS task. For vehicle detection, we \changed{used} the classes ``vehicle'', ``road'', ``environment'' and ``sky'', whereas, for lane-keeping, we considered the classes ``road'' (including central and outer lanes) and ``background''.
Subsequently, we employ the trained SegFormer model to assess the semantic differences between the original image and the translated version, following a similar approach as for the SSS metric. 

To offer varying levels of fine-tuning, we compute semantic difference values both (i)~across all \changed{relevant} classes (Targeted Semantic Segmentation \metric) and (ii)~only for the most relevant class of each ADS task---i.e., ``vehicle'' for vehicle detection and ``road'' for lane keeping---(One Class - Targeted Semantic Segmentation \onemetric).
After computing the \metric and \onemetric scores, we computed the correlation between both fine-tuned image metric values and the ADS behavior metrics as done for RQ\textsubscript{2}.

\begin{table}[t]
\caption{Distribution-level metrics relative error correlations.}\label{tab:rq2_dist}
\setlength{\tabcolsep}{2.4pt}
\renewcommand{\arraystretch}{1.1}

\begin{tabular}{lrrrrrrr}
\toprule

& \multicolumn{3}{c}{\bf Vehicle Detection} && \multicolumn{3}{c}{\bf Lane Keeping} \\
\cmidrule{2-4}
\cmidrule{6-8}
& {Prediction}  & {Confidence} & {Attention} && {Prediction}  & {Confidence} & {Attention} \\

\midrule
\changed{IS} & \textbf{0.41} &  0.37 & 0.41 && 0.14  & 0.72 & 0.65\\
\changed{FID} & 0.24 &  0.21 & 0.31 && 0.64  & \textbf{0.86} & \textbf{0.78}\\
\changed{KID} & 0.54 &  0.64 & \textbf{0.86} && 0.54 &  \textbf{0.74} & 0.60 \\

\bottomrule
\end{tabular}
\end{table}

\begin{table*}
\caption{Single-image metrics correlations. \textbf{Bold values} indicate statistical significance (\textit{p}-value $< 0.05$), \badcorrelation{red values} indicate wrong correlation direction, and \lowcorrelation{grey values} indicate correlations lower than $0.10$. $\pmb{\downarrow}$/$\pmb{\uparrow}=$ lower/higher values are preferable.}\label{tab:rq2_rq3_a}
\centering

\footnotesize

\setlength{\tabcolsep}{4.72pt}
\renewcommand{\arraystretch}{1.1}

\begin{tabular}{@{}lrrrrrrrrrrrrrrrrrrr@{}}

\toprule



&& \multicolumn{6}{c}{\bf Relative Prediction Error} & \multicolumn{6}{c}{\bf Relative Confidence Error} & \multicolumn{6}{c}{\bf Relative Attention Error} \\

\cmidrule(r){3-8}
\cmidrule(r){9-14}
\cmidrule(r){15-20}

&& \multicolumn{1}{c}{C1} & \multicolumn{1}{c}{C2} & \multicolumn{1}{c}{C3} & \multicolumn{1}{c}{P1} & \multicolumn{1}{c}{P2} & \multicolumn{1}{c}{P3} & \multicolumn{1}{c}{C1} & \multicolumn{1}{c}{C2} & \multicolumn{1}{c}{C3} & \multicolumn{1}{c}{P1} & \multicolumn{1}{c}{P2} & \multicolumn{1}{c}{P3} & \multicolumn{1}{c}{C1} & \multicolumn{1}{c}{C2} & \multicolumn{1}{c}{C3} & \multicolumn{1}{c}{P1} & \multicolumn{1}{c}{P2} & \multicolumn{1}{c}{P3} \\
\midrule
&\multicolumn{1}{l}{\textbf{RQ\textsubscript{2}}}\\
\parbox[t]{0.5mm}{\multirow{12}{*}{\rotatebox[origin=c]{90}{\sc Vehicle detection}}} & SSIM $\pmb{\uparrow}$ & \lowcorrelation{-0.03} & \lowcorrelation{\textbf{0.06}} & \lowcorrelation{-0.02} & \badcorrelation{\textbf{0.18}} & \badcorrelation{\textbf{0.21}} & \badcorrelation{\textbf{0.26}} & \badcorrelation{\textbf{0.16}} & \badcorrelation{\textbf{0.11}} & \badcorrelation{\textbf{0.20}} & \lowcorrelation{\textbf{0.06}} & \lowcorrelation{0.05} & \lowcorrelation{\textbf{0.08}} & \lowcorrelation{\textbf{-0.09}} & \lowcorrelation{-0.04} & \lowcorrelation{-0.02} & \lowcorrelation{\textbf{-0.09}} & \textbf{-0.14} & \textbf{-0.26} \\
&PSNR $\pmb{\uparrow}$& \lowcorrelation{\textbf{0.08}} & \lowcorrelation{\textbf{0.07}} & \lowcorrelation{\textbf{-0.09}} & \badcorrelation{\textbf{0.30}} & \badcorrelation{\textbf{0.32}} & \badcorrelation{\textbf{0.35}} & \badcorrelation{\textbf{0.18}} & \lowcorrelation{0.04} & \badcorrelation{\textbf{0.14}} & \lowcorrelation{-0.01} & \lowcorrelation{0.01} & \badcorrelation{\textbf{0.12}} & \textbf{-0.11} & \lowcorrelation{-0.01} & \lowcorrelation{-0.01} & \lowcorrelation{-0.01} & \lowcorrelation{\textbf{-0.08}} & \textbf{-0.16} \\
&MSE $\pmb{\downarrow}$& \lowcorrelation{\textbf{-0.08}} & \lowcorrelation{\textbf{-0.08}} & \lowcorrelation{\textbf{0.09}} & \badcorrelation{\textbf{-0.30}} & \badcorrelation{\textbf{-0.31}} & \badcorrelation{\textbf{-0.35}} & \badcorrelation{\textbf{-0.18}} & \lowcorrelation{-0.05} & \badcorrelation{\textbf{-0.14}} & \lowcorrelation{0.01} & \lowcorrelation{-0.01} & \badcorrelation{\textbf{-0.12}} & \textbf{0.11} & \lowcorrelation{0.02} & \lowcorrelation{0.00} & \lowcorrelation{0.00} & \lowcorrelation{\textbf{0.08}} & \textbf{0.16} \\
&CS $\pmb{\uparrow}$& \lowcorrelation{-0.03} & \lowcorrelation{0.04} & \lowcorrelation{\textbf{-0.06}} & \textbf{-0.21} & \textbf{-0.14} & \badcorrelation{\textbf{0.18}} & \badcorrelation{\textbf{0.18}} & \badcorrelation{\textbf{0.13}} & \lowcorrelation{0.04} & \lowcorrelation{-0.01} & \lowcorrelation{\textbf{0.09}} & \badcorrelation{\textbf{0.20}} & \badcorrelation{\textbf{0.10}} & \lowcorrelation{0.03} & \lowcorrelation{0.01} & \badcorrelation{\textbf{0.32}} & \badcorrelation{\textbf{0.20}} & \textbf{-0.11} \\
&TSI $\pmb{\downarrow}$& \lowcorrelation{\textbf{0.07}} & \lowcorrelation{0.00} & \badcorrelation{\textbf{-0.13}} & \lowcorrelation{\textbf{-0.08}} & \badcorrelation{\textbf{-0.13}} & \badcorrelation{\textbf{-0.16}} & \lowcorrelation{0.05} & \lowcorrelation{0.01} & \badcorrelation{\textbf{-0.12}} & \lowcorrelation{0.04} & \lowcorrelation{-0.05} & \lowcorrelation{-0.01} & \badcorrelation{\textbf{-0.20}} & \textbf{0.15} & \textbf{0.13} & \lowcorrelation{\textbf{-0.09}} & \textbf{0.15} & \textbf{0.37} \\
&WD $\pmb{\downarrow}$& \lowcorrelation{\textbf{-0.08}} & \lowcorrelation{0.00} & \lowcorrelation{0.04} & \lowcorrelation{0.06} & \lowcorrelation{\textbf{0.09}} & \lowcorrelation{\textbf{0.09}} & \lowcorrelation{0.05} & \lowcorrelation{\textbf{0.07}} & \textbf{0.31} & \textbf{0.17} & \textbf{0.11} & \textbf{0.15} & \badcorrelation{\textbf{-0.10}} & \lowcorrelation{0.00} & \lowcorrelation{-0.06} & \badcorrelation{\textbf{-0.30}} & \badcorrelation{\textbf{-0.28}} & \badcorrelation{\textbf{-0.35}} \\
&KL $\pmb{\downarrow}$& \lowcorrelation{0.03} & \badcorrelation{\textbf{-0.13}} & \textbf{0.14} & \badcorrelation{\textbf{-0.11}} & \lowcorrelation{0.04} & \textbf{0.12} & \lowcorrelation{\textbf{0.07}} & \badcorrelation{\textbf{-0.11}} & \textbf{0.13} & \lowcorrelation{0.04} & \textbf{0.18} & \textbf{0.14} & \lowcorrelation{\textbf{-0.08}} & \textbf{0.15} & \badcorrelation{\textbf{-0.11}} & \textbf{0.18} & \lowcorrelation{\textbf{-0.08}} & \badcorrelation{\textbf{-0.18}} \\
&HistI $\pmb{\uparrow}$& \lowcorrelation{\textbf{0.09}} & \lowcorrelation{0.05} & \lowcorrelation{\textbf{-0.06}} & \badcorrelation{\textbf{0.23}} & \badcorrelation{\textbf{0.17}} & \badcorrelation{\textbf{0.31}} & \lowcorrelation{0.05} & \badcorrelation{\textbf{0.10}} & \lowcorrelation{-0.06} & \lowcorrelation{-0.00} & \lowcorrelation{\textbf{-0.09}} & \lowcorrelation{\textbf{0.06}} & \lowcorrelation{-0.01} & \textbf{-0.12} & \lowcorrelation{\textbf{0.07}} & \lowcorrelation{0.02} & \lowcorrelation{0.03} & \textbf{-0.27} \\
&CPL $\pmb{\downarrow}$& \lowcorrelation{0.00} & \lowcorrelation{0.01} & \lowcorrelation{0.02} & \textbf{0.29} & \textbf{0.20} & \textbf{0.16} & \textbf{0.23} & \textbf{0.19} & \textbf{0.11} & \lowcorrelation{0.00} & \textbf{0.15} & \lowcorrelation{\textbf{0.08}} & \badcorrelation{\textbf{-0.10}} & \lowcorrelation{-0.01} & \lowcorrelation{-0.01} & \textbf{0.17} & \textbf{0.26} & \textbf{0.36} \\
&SSS $\pmb{\downarrow}$& \lowcorrelation{-0.02} & \lowcorrelation{-0.01} & \lowcorrelation{0.01} & \textbf{0.23} & \textbf{0.19} & \textbf{0.10} & \textbf{0.21} & \textbf{0.17} & \textbf{0.13} & \lowcorrelation{0.01} & \textbf{0.18} & \lowcorrelation{\textbf{0.09}} & \badcorrelation{\textbf{-0.12}} & \lowcorrelation{-0.04} & \lowcorrelation{-0.03} & \textbf{0.21} & \textbf{0.23} & \textbf{0.34} \\
&\multicolumn{1}{l}{\textbf{RQ\textsubscript{3}}}\\
&\metric $\pmb{\downarrow}$& \lowcorrelation{0.05} & \lowcorrelation{\textbf{-0.08}} & \lowcorrelation{\textbf{-0.09}} & \textbf{0.26} & \textbf{0.14} & \lowcorrelation{0.01} & \lowcorrelation{-0.05} & \lowcorrelation{-0.02} & \badcorrelation{\textbf{-0.11}} & \lowcorrelation{-0.04} & \lowcorrelation{0.05} & \lowcorrelation{-0.03} & \lowcorrelation{0.04} & \textbf{0.19} & \textbf{0.13} & \textbf{0.11} & \textbf{0.21} & \textbf{0.36} \\
&\onemetric $\pmb{\downarrow}$& \textbf{0.33} & \textbf{0.21} & \textbf{0.15} & \textbf{0.57} & \textbf{0.55} & \textbf{0.49} & \textbf{0.13} & \textbf{0.19} & \textbf{0.21} & \lowcorrelation{\textbf{0.08}} & \textbf{0.10} & \textbf{0.12} & \badcorrelation{\textbf{-0.10}} & \lowcorrelation{\textbf{-0.08}} & \lowcorrelation{\textbf{-0.09}} & \badcorrelation{\textbf{-0.24}} & \badcorrelation{\textbf{-0.19}} & \badcorrelation{\textbf{-0.24}}\\
\midrule
&\multicolumn{1}{l}{\textbf{RQ\textsubscript{2}}}\\
\parbox[t]{0.5mm}{\multirow{12}{*}{\rotatebox[origin=c]{90}{\sc Lane keeping}}} &SSIM $\pmb{\uparrow}$& \textbf{-0.31} & \textbf{-0.28} & \textbf{-0.39} & \lowcorrelation{-0.09} & \lowcorrelation{-0.03} & \textbf{-0.15} & -0.14 & -0.13 & \lowcorrelation{-0.05} & -0.12 & -0.07 & -0.13 & \textbf{-0.34} & \textbf{-0.37} & \textbf{-0.20} & \lowcorrelation{-0.03} & \lowcorrelation{0.05} & \badcorrelation{\textbf{0.35}} \\
&PSNR $\pmb{\uparrow}$& \textbf{-0.33} & \textbf{-0.34} & \textbf{-0.45} & -0.12 & \lowcorrelation{-0.08} & -0.12 & -0.11 & \lowcorrelation{-0.02} & \lowcorrelation{-0.08} & -0.13 & \lowcorrelation{-0.01} & -0.11 & \textbf{-0.38} & \textbf{-0.36} & \textbf{-0.31} & \lowcorrelation{-0.02} & -0.10 & \badcorrelation{\textbf{0.18}} \\
&MSE $\pmb{\downarrow}$& \textbf{0.34} & \textbf{0.34} & \textbf{0.45} & 0.12 & \lowcorrelation{0.07} & 0.12 & 0.11 & \lowcorrelation{0.02} & \lowcorrelation{0.08} & 0.13 & \lowcorrelation{0.01} & 0.11 & \textbf{0.38} & \textbf{0.36} & \textbf{0.31} & \lowcorrelation{0.01} & 0.10 & \badcorrelation{\textbf{-0.17}} \\
&CS $\pmb{\uparrow}$& \lowcorrelation{0.03} & \lowcorrelation{0.00} & \lowcorrelation{-0.07} & \lowcorrelation{-0.06} & \textbf{-0.14} & \lowcorrelation{0.02} & \lowcorrelation{-0.04} & -0.10 & \lowcorrelation{-0.05} & \lowcorrelation{-0.03} & -0.13 & \lowcorrelation{0.02} & -0.14 & \lowcorrelation{-0.01} & \lowcorrelation{-0.06} & -0.11 & \badcorrelation{0.14} & \lowcorrelation{-0.00} \\
&TSI $\pmb{\downarrow}$& \badcorrelation{\textbf{-0.15}} & \badcorrelation{-0.12} & \lowcorrelation{-0.06} & \badcorrelation{\textbf{-0.26}} & \badcorrelation{\textbf{-0.18}} & \lowcorrelation{0.00} & 0.10 & 0.11 & \textbf{0.20} & \badcorrelation{-0.10} & \badcorrelation{-0.14} & \lowcorrelation{-0.07} & \lowcorrelation{-0.04} & \lowcorrelation{-0.07} & \badcorrelation{\textbf{-0.18}} & \badcorrelation{\textbf{-0.18}} & \badcorrelation{\textbf{-0.16}} & \badcorrelation{\textbf{-0.16}} \\
&WD $\pmb{\downarrow}$& \lowcorrelation{0.07} & \lowcorrelation{0.02} & \badcorrelation{\textbf{-0.15}} & \lowcorrelation{-0.07} & \lowcorrelation{-0.07} & \badcorrelation{\textbf{-0.16}} & 0.10 & \lowcorrelation{0.07} & \lowcorrelation{0.07} & 0.13 & 0.10 & \lowcorrelation{0.07} & \lowcorrelation{-0.05} & \lowcorrelation{0.00} & \lowcorrelation{-0.08} & \badcorrelation{\textbf{-0.19}} & \lowcorrelation{0.03} & \textbf{0.22} \\
&KL $\pmb{\downarrow}$& \textbf{0.25} & 0.10 & 0.10 & \lowcorrelation{-0.05} & \lowcorrelation{0.00} & \lowcorrelation{-0.05} & \textbf{0.21} & 0.10 & \lowcorrelation{0.07} & 0.11 & \textbf{0.18} & \textbf{0.19} & \textbf{0.27} & \textbf{0.29} & \textbf{0.37} & \lowcorrelation{-0.01} & \lowcorrelation{-0.09} & \badcorrelation{-0.12} \\
&HistI $\pmb{\uparrow}$& \textbf{-0.15} & \lowcorrelation{0.01} & \textbf{-0.24} & \badcorrelation{\textbf{0.21}} & \lowcorrelation{0.03} & \textbf{-0.27} & \badcorrelation{0.11} & \lowcorrelation{0.08} & \badcorrelation{0.13} & \badcorrelation{0.11} & \badcorrelation{\textbf{0.20}} & \lowcorrelation{-0.01} & -0.13 & \badcorrelation{0.10} & \textbf{-0.17} & \lowcorrelation{0.05} & \textbf{-0.37} & \lowcorrelation{0.03} \\
&CPL $\pmb{\downarrow}$& \textbf{0.15} & \textbf{0.28} & \textbf{0.30} & \textbf{0.22} & \textbf{0.17} & \lowcorrelation{0.02} & \textbf{0.18} & \textbf{0.25} & 0.12 & \textbf{0.15} & \textbf{0.19} & \textbf{0.30} & \lowcorrelation{-0.05} & \badcorrelation{-0.13} & \badcorrelation{-0.13} & \lowcorrelation{0.01} & 0.14 & \textbf{0.16} \\
&SSS $\pmb{\downarrow}$& \textbf{0.21} & \textbf{0.18} & \textbf{0.25} & 0.11 & \badcorrelation{-0.10} & \lowcorrelation{-0.05} & \textbf{0.23} & \textbf{0.30} & 0.12 & \textbf{0.21} & \textbf{0.19} & \textbf{0.26} & \lowcorrelation{-0.02} & \lowcorrelation{-0.05} & \badcorrelation{-0.11} & \lowcorrelation{-0.02} & \textbf{0.26} & \lowcorrelation{0.07} \\
&\multicolumn{1}{l}{\textbf{RQ\textsubscript{3}}}\\
&\metric $\pmb{\downarrow}$& \textbf{0.15} & \textbf{0.16} & \textbf{0.16} & \textbf{0.17} & 0.13 & \textbf{0.20} & \textbf{0.15} & \textbf{0.16} & \textbf{0.16} & \textbf{0.17} & 0.13 & \textbf{0.20} & \lowcorrelation{0.07} & \lowcorrelation{-0.01} & \badcorrelation{\textbf{-0.17}} & \lowcorrelation{0.08} & \badcorrelation{-0.10} & \badcorrelation{\textbf{-0.48}} \\
&\onemetric $\pmb{\downarrow}$& \textbf{0.39} & \textbf{0.35} & \textbf{0.44} & \textbf{0.42} & \textbf{0.31} & \textbf{0.38} & \textbf{0.14} & \textbf{0.15} & \textbf{0.15} & \textbf{0.18} & 0.13 & \textbf{0.19} & \lowcorrelation{0.06} & \lowcorrelation{-0.04} & \badcorrelation{\textbf{-0.16}} & 0.10 & \lowcorrelation{-0.08} & \badcorrelation{\textbf{-0.45}}\\
\bottomrule

\end{tabular}
\end{table*}

\subsection{Results}

\head{RQ\textsubscript{1} (Transferability)}
\autoref{fig:rq1}~(top) shows the results for the vehicle detection ADS, whereas \autoref{fig:rq1}~(bottom) shows the results for the lane-keeping ADS. For both ADS tasks, each row displays several boxplots representing the ADS behavior \changed{deviation from the real counterpart} (prediction error, confidence, and attention error) on simulated data (Sim) and I2I-generated data for the I2I models under test at various epochs (C1, C2, C3, P1, P2, P2, see \autoref{tab:table_model_epochs_all}). 

Considering the relative prediction and attention errors, C3 has a good performance in mitigating the sim2real gap in the vehicle detection task, whereas, for the lane-keeping task, our results show both I2I models mitigate the sim2real gap.
Our results reveal that the relative confidence error may not be useful for sim2real gap mitigation assessment, as none of the models of either domain show an improvement over the simulated inputs.
We assess the statistical significance of the differences between real-world and simulator/I2I errors using a paired t-test~\cite{t-test} (with $\alpha = 0.05$), the magnitude of the differences using the Cohen's $d$ effect size~\cite{cohen1988statistical}, and the statistical power with a Monte Carlo power analysis~\cite{mcpower} with 80\% power target as our data is not normally distributed.

In the vehicle detection task, only C3 has a distribution of prediction errors for which it was the same for real-world and for I2I-generated images (i.e., $p$-value $\geq$ 0.05, and high statistical power). This indicates that a high-quality CycleGAN architecture can generate sim2real gap mitigating inputs for which the ADS outputs predictions similar to those on real-world data. 
Differently, none of the domains showed statistically significant differences for confidence and attention errors ($p$-value $<$ 0.05). These findings indicate that such ADS behavioral metrics are less adequate to analyze the ADS transferability across virtual and real-world environments at the model level.
In the lane-keeping task, C2, C3, and P2 have a distribution of prediction and attention errors that were the same for the real world and for I2I-generated images (i.e., $p$-value $\geq$ 0.05, and high statistical power). Regarding confidence, our analysis does not identify any distribution as statistically close. 

\begin{tcolorbox}
\textbf{RQ\textsubscript{1} (Transferability)} The effectiveness of I2I models for mitigating the sim2real gap varies across different tasks. For lane keeping, all I2I models mitigate the sim2real gap in terms of prediction and attention errors; for vehicle detection, only a high-quality CycleGAN achieves this outcome.
\end{tcolorbox}

\head{RQ\textsubscript{2} (Correlation)}
Concerning distribution-level metrics, \autoref{tab:rq2_dist} presents the Pearson's correlation coefficients between the distribution-level I2I models evaluation metrics (first column) and the ADS evaluation metrics (remaining columns). Also in this case results are aggregated by ADS task. Bold values indicate statistical significance (\textit{p}-value $< 0.05$). 

Our results indicate good correlations between all distribution-level metrics and ADS metrics in both ADS tasks, except for IS and prediction error in the lane-keeping task. Specifically, in the vehicle detection task, the IS metric correlates with statistical significance with relative prediction error, and KID does so with attention error. In contrast, in the lane-keeping task, FID demonstrates a significant correlation with both Uncertainty and attention error, while KID does so with uncertainty.

\changed{Concerning image-level metrics, \autoref{tab:rq2_rq3_a} present the Pearson's correlation coefficients between the single-image metrics (rows) and the ADS evaluation metrics (columns) on each generated domain (C1, C2, C3, P1, P2, P3) for both. Results are aggregated by ADS task (Vehicle detection and Lane keeping).}
Overall, our results show correlation ranges $\mp0.45$; for the sake of space, we limit our analysis to metrics having absolute correlation scores higher than 0.10, with statistical significance.
For the vehicle detection task, CPL and SSS are the only metrics that exhibit a statistically significant and consistent correlation, without directional disagreement, with the ADS relative prediction errors. This trend is observed exclusively in pix2pix-based models (P1, P2, and P3). On the other hand, in the lane-keeping task, SSIM, PSNR, MSE, and SSS exhibit this behavior for CycleGAN-based models, while CPL achieves it in both tasks.
In terms of relative confidence error, WD, CPL, and SSS exhibit statistically significant correlations without directional discrepancies in the vehicle detection task, while, for the lane-keeping task, this is observed only with CPL and SSS.
Lastly, concerning Attention error, SSIM, PSNR, and MSE show a correlation with some models in the vehicle detection task, whereas none of the metrics do so in the lane-keeping task.

\begin{tcolorbox}
\textbf{RQ\textsubscript{2} (Correlation):} All distribution-level metrics correlate well with \changed{at least one} ADS behavior metric. The correlation does not transfer across ADS tasks. Single-image metrics, on the contrary, are highly inconsistent both across ADS tasks and I2I models, making them less suitable for sim2real ADS assessment.
\end{tcolorbox}
\begin{figure}
\centering
\includegraphics[width=\linewidth]{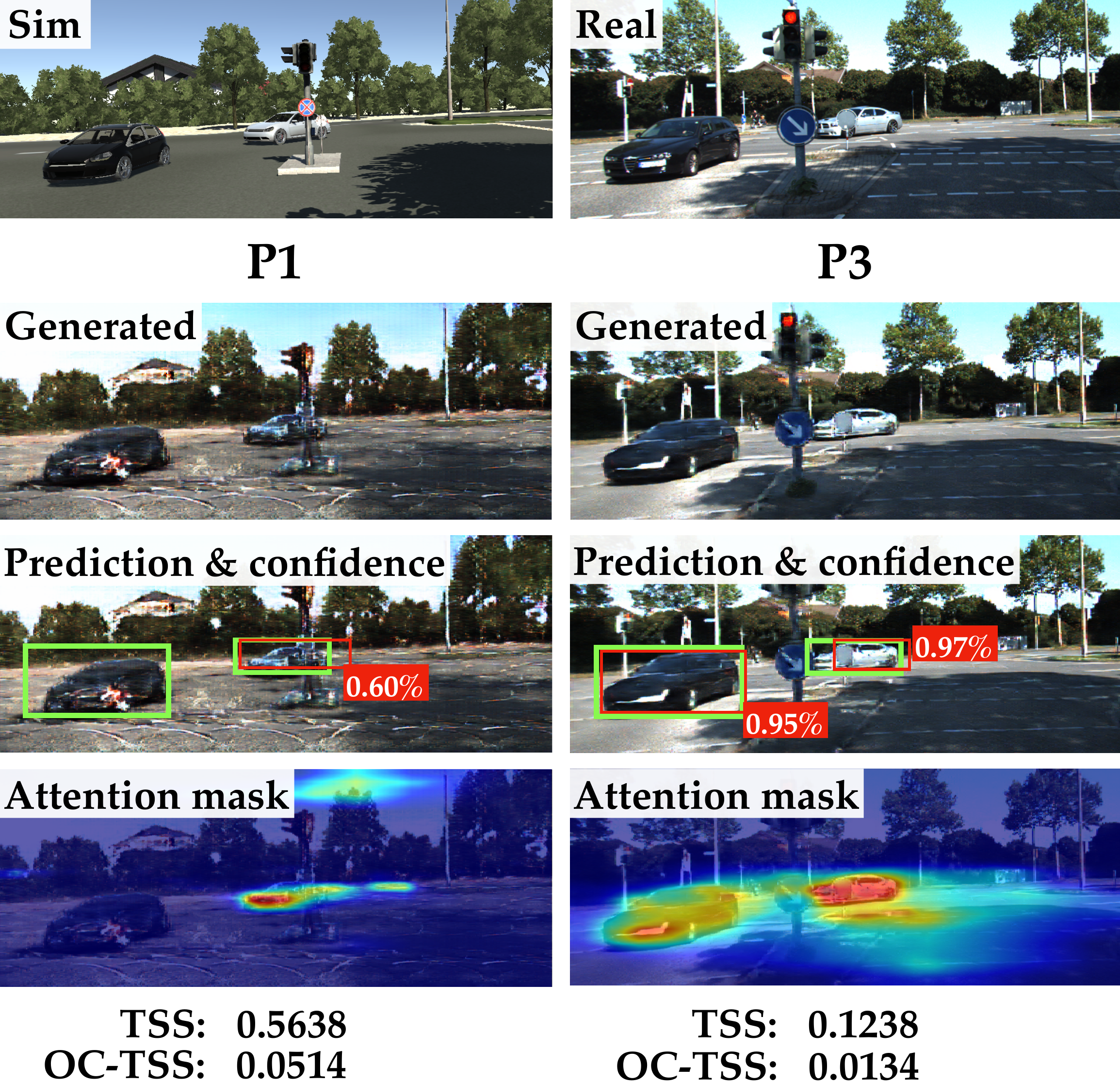}

\caption{Translation quality comparison: P1 (low-quality pix2pix) and P3 (high-quality pix2pix). Rows depict simulated/reference images, neural-generated images by P1 and P3, ground truth and vehicle detection ground truth and output (green and red boxes) with confidence (in percentage), ADS attention mask, and TSS/OC-TSS values---larger values of the metrics highlight the error between the semantic representation of the original and generated images, thus characterizing neural translation effectiveness.}
\label{fig:translation_quality}
\vspace{-1.5em}
\end{figure}
\head{RQ\textsubscript{3} (Fine-tuning)}
The bottom part of \autoref{tab:rq2_rq3_a}, shows, for each task, the Pearson's correlation coefficients between the \metric and \onemetric metrics (first column) and the ADS evaluation metrics (remaining columns). 

The metric \metric, which utilizes all segmentation classes to calculate image difference, exhibits correlation scores with the ADS behavior in line with existing metrics (RQ\textsubscript{2}) for both ADS tasks (values up to 0.36), but notably does not show directional correlation disagreement for prediction error and attention error in the vehicle detection task and for prediction error and confidence error in the lane-keeping task.
Across ADS tasks, \metric correlates negatively with the attention error for P1 and P3 in the lane-keeping task, whereas it correlates positively in the vehicle detection task for all models.  

Differently, \onemetric is the only metric in our study that shows stable correlations across ADS tasks. Notably, it also achieves the highest correlation scores for P1, P2, and P3 (vehicle detection). It shows positive correlations with relative prediction errors in both ADS tasks, for both pix2pix and CycleGAN models.

Finally, it is worth noting that when it comes to the vehicle detection task, \onemetric stands out as the only metric that consistently exhibits a statistically significant positive correlation with all evaluation metrics across various generative domains (specifically, pix2pix and CycleGAN). The only exception to this trend is the attention scores, which display a negative correlation, in contrast with the expected metric direction.

\begin{tcolorbox}
\textbf{RQ\textsubscript{3} (Fine-tuning):} In both evaluated task domains, \metric and \onemetric provide valuable measures to evaluate the quality of I2I translation models, in terms of relative prediction and confidence error, making them suitable for sim2real ADS assessment.
\end{tcolorbox}

\autoref{fig:translation_quality} shows some meaningful examples from our study. 
We present an example of low-quality and high-quality I2I translations produced by pix2pix models P1 and P3.
For each model output (second row), we provide the predictions of the vehicle detection model, including confidence scores (third row), and the associated attention maps (fourth row).
Below each column, we present the \metric and \onemetric metrics that we have developed to address RQ\textsubscript{3}. These metrics indicate a significant decline in quality, with a five-fold difference similar to the decrease in ADS performance, when comparing the visually inferior translation to the superior one. This drop in quality is evident in incomplete vehicle classification, increased false negatives, reduced confidence in predictions, and less focused attention.

\subsection{Threats to Validity}

\subsubsection{Internal validity}
In the absence of a consensus on appropriate metrics for evaluating I2I models, we encountered challenges in determining when to halt the training of our I2I translation architectures. Consequently, we extended the training of our generators until the generated images reached a visually high-quality standard. To ensure impartiality in our approach, we trained each architecture for an equal duration in both task domains and recorded model checkpoints. From these checkpoints, we arbitrarily selected three training epochs (10-50-90\%) for reporting.

For both ADS task evaluations, the requirement was to have pairs of simulated and real images. Unfortunately, to our knowledge, the only existing lane-keeping dataset~\cite{2020-Haq-ICST}, paired images based on steering angles rather than sim-to-real image similarity.
Consequently, we faced a shortage of suitable resources and had to resort to manual mapping of a simulated and real dataset~\cite{2023-Stocco-TSE}. These datasets depicted the same scenes, but they were not originally designed to provide a one-to-one correspondence between simulated and real images. This manual mapping process may have introduced minor content inconsistencies due to its non-standardized nature.

In the case of the \davetwo lane-keeping model used in one of our ADS domains, our results are contingent on the training quality of this model. We adhered to best training practices to ensure that the results reflected the behavior of a well-performing model. However, it is important to note that results may vary if the model is trained with different parameters.

\subsubsection{External Validity}
Regarding external validity, it is worth acknowledging that our study is limited by the number of ADS systems and I2I models considered, due to the experiment requirements \changednew{(overall, we computed 340$k$ scores $=$ 6 I2I models $\times$ 15 I2I metrics $\times$ 3 ADS metrics $\times$ 1,261 images).} This limitation can impact the generalizability of our findings.

\subsubsection{Reproducibility} 
The entire pipeline used to obtain the results discussed in this empirical study, including I2I model training, segmentation models, metric calculations, and comparisons, is accessible and can be reproduced~\cite{tool}.
\section{Related work}\label{sec:related-work}

Existing research in the field of autonomous driving has emphasized the need for real-world testing of ADS. Simulation platforms, while valuable in the initial stages of testing, are characterized by limitations in faithfully replicating the complex characteristics of real-world environments~\cite{AfzalSimulation21,10.1145/3368089.3409743}. For a comprehensive exploration of the reality gap problem in ADS, we recommend referring to the existing literature~\cite{hu2023sim2real}.

Sim2real gap mitigation techniques are highly diverse, including digital twins~\cite{schwarz2022role}, transfer learning~\cite{kalapos2020sim}, adversarial testing~\cite{deepbillboard}, multi-simulation environments~\cite{biagiola2023better}, domain randomization~\cite{domain_rando}, and hardware-in-the-loop testing~\cite{fathy2006review}.

In this paper, we focus on reality gap mitigation at an input level through the utilization of GANs for real-world like ADS inputs. 
In the context of model-level testing~\cite{2023-Stocco-EMSE,2020-Haq-ICST}, researchers have employed GAN-generated driving scenarios to evaluate the performance of these systems. DeepRoad~\cite{deeproad} leverages UNIT~\cite{unit} to generate highly realistic paired driving scenes for testing self-driving cars. 

Concerning system-level testing, SilGAN~\cite{DBLP:journals/corr/abs-2107-07364} uses GANs to create driving maneuvers for software-in-the-loop testing. Kong et al.~\cite{physgan} developed realistic adversarial billboards embedded in real-world images to potentially confuse autonomous vehicles. DeepBillboard~\cite{deepbillboard} employs both digital and physical adversarial perturbation techniques to influence offline steering. SurfelGAN~\cite{DBLP:journals/corr/abs-2005-03844} is a technique developed by Waymo to generate realistic sensor data for autonomous driving simulations without the need for manually creating virtual environments and objects.
A recent study~\cite{2023-Stocco-TSE} uses CycleGAN to generate real-world-like images from a driving simulator to train a telemetry predictor. Our findings reveal that the efficacy of CycleGAN can vary depending on the specific tasks, including factors related to ADS outputs.

To improve the capabilities of GANs and address image quality issues, numerous studies have examined different GAN evaluation metrics. However, there is no clear consensus on which metrics are most suitable for assessing GAN performance~\cite{DBLP:journals/corr/abs-1802-03446,i2i-trans-metrics}, even less when considering the sim2real gap in ADS testing.

Some studies have empirically validated GANs but have primarily focused on analyzing the resulting distributions without specific task-oriented evaluations~\cite{empirical_2}, or they have used general-purpose ImageNet models~\cite{huang2018an}.
In our work, instead, we evaluate the available metrics in correlation with the ADS performance under test and propose a task-specific metric that captures the behavior differences caused by synthetic inputs generated by different GAN-based I2I translation models in the context of image-gap mitigation.

To the best of our knowledge, our study is the first that targets the reality gap mitigation for ADS inputs using generative I2I translation models and evaluates different metrics to quantify the gap between virtual and real imagery data for ADS testing. 

\section{Conclusions and Future Work}\label{sec:conclusions}

This work addresses the sim2real gap for vision-based ADS by employing data-driven neural I2I translation models to transform simulated driving scene images into real-world driving counterparts.
I2I models can generate realistic images but have limitations, making it essential to have reliable metrics that can assess whether the inputs generated by such techniques are sim2real gap mitigating.

To this aim, this study contributes by empirically evaluating 13 existing metrics and two I2I models in addressing the sim-to-real gap in ADS testing. Our study highlights the limited transferability of I2I models across ADS domains and it reveals that these metrics lack consistency across tasks. This issue is addressed by introducing a task-specific perception metric, which is shown to have higher transferability. 

Concerning future work, we plan to employ the best metrics from our study to develop data-driven techniques for sim2real-mitigating simulation-based testing of ADS. Additionally, these metrics can be utilized in combination with input validators~\cite{2024-Grewal-ICST,2022-Stocco-ASE,2020-Stocco-ICSE,2020-Stocco-GAUSS,2021-Stocco-JSEP} to filter out low-quality real-world data that are not expected to be supported by the ADS. 
Recent propositions showed the ability of diffusion models and neural radiance fields to generate high-quality images, which we plan to study as well in our future work. Additionally, we plan to expand our study to other I2I models and ADS tasks, including evaluations at a system level employing driving simulation platforms.

\section*{Acknowledgements}
\addcontentsline{toc}{section}{Acknowledgements}
This research was funded by the Bavarian Ministry of Economic Affairs, Regional Development and Energy.

\balance
\bibliographystyle{IEEEtran}
\bibliography{paper}

\begin{thebibliography}{10}
\providecommand{\url}[1]{#1}
\csname url@samestyle\endcsname
\providecommand{\newblock}{\relax}
\providecommand{\bibinfo}[2]{#2}
\providecommand{\BIBentrySTDinterwordspacing}{\spaceskip=0pt\relax}
\providecommand{\BIBentryALTinterwordstretchfactor}{4}
\providecommand{\BIBentryALTinterwordspacing}{\spaceskip=\fontdimen2\font plus
\BIBentryALTinterwordstretchfactor\fontdimen3\font minus
  \fontdimen4\font\relax}
\providecommand{\BIBforeignlanguage}[2]{{%
\expandafter\ifx\csname l@#1\endcsname\relax
\typeout{** WARNING: IEEEtran.bst: No hyphenation pattern has been}%
\typeout{** loaded for the language `#1'. Using the pattern for}%
\typeout{** the default language instead.}%
\else
\language=\csname l@#1\endcsname
\fi
#2}}
\providecommand{\BIBdecl}{\relax}
\BIBdecl

\bibitem{yurtsever2020survey}
E.~Yurtsever, J.~Lambert, A.~Carballo, and K.~Takeda, ``A survey of autonomous
  driving: Common practices and emerging technologies,'' \emph{IEEE access},
  vol.~8, pp. 58\,443--58\,469, 2020.

\bibitem{chen2023endtoend}
L.~Chen, P.~Wu, K.~Chitta, B.~Jaeger, A.~Geiger, and H.~Li, ``End-to-end
  autonomous driving: Challenges and frontiers,'' 2023.

\bibitem{Cerf:2018:CSC:3181977.3177753}
V.~G. Cerf, ``A comprehensive self-driving car test,'' \emph{Commun. ACM},
  vol.~61, no.~2, pp. 7--7, Jan. 2018.

\bibitem{zhong2021survey}
Z.~Zhong, Y.~Tang, Y.~Zhou, V.~de~Oliveira~Neves, Y.~Liu, and B.~Ray, ``A
  survey on scenario-based testing for automated driving systems in
  high-fidelity simulation,'' 2021.

\bibitem{2020-Riccio-EMSE}
V.~Riccio, G.~Jahangirova, A.~Stocco, N.~Humbatova, M.~Weiss, and P.~Tonella,
  ``{Testing Machine Learning based Systems: A Systematic Mapping},''
  \emph{Empirical Software Engineering}, 2020.

\bibitem{AfzalSimulation21}
A.~Afzal, D.~S. Katz, C.~Le~Goues, and C.~S. Timperley, ``{Simulation for
  Robotics Test Automation: Developer Perspectives},'' in \emph{2021 14th IEEE
  Conference on Software Testing, Verification and Validation (ICST)}, 2021,
  pp. 263--274.

\bibitem{ADS_testing_survey_2022}
S.~Tang, Z.~Zhang, Y.~Zhang, J.~Zhou, Y.~Guo, S.~Liu, S.~Guo, Y.-F. Li, L.~Ma,
  Y.~Xue, and Y.~Liu, ``A survey on automated driving system testing:
  Landscapes and trends,'' 2022.

\bibitem{Kim2021_DriveGAN}
S.~W. Kim, , J.~Philion, A.~Torralba, and S.~Fidler, ``{DriveGAN: Towards a
  Controllable High-Quality Neural Simulation},'' in \emph{IEEE Conference on
  Computer Vision and Pattern Recognition (CVPR)}, Jun. 2021.

\bibitem{DBLP:journals/corr/abs-1802-03446}
A.~Borji, ``{Pros and Cons of GAN Evaluation Measures},'' \emph{CoRR}, vol.
  abs/1802.03446, 2018.

\bibitem{i2i-trans-metrics}
Y.~Pang, J.~Lin, T.~Qin, and Z.~Chen, ``Image-to-image translation: Methods and
  applications,'' \emph{IEEE Transactions on Multimedia}, vol.~24, pp.
  3859--3881, 2022.

\bibitem{NIPS2014_5423}
I.~Goodfellow, J.~Pouget-Abadie, M.~Mirza, B.~Xu, D.~Warde-Farley, S.~Ozair,
  A.~Courville, and Y.~Bengio, ``Generative adversarial nets,'' in
  \emph{Advances in Neural Information Processing Systems 27}, Z.~Ghahramani,
  M.~Welling, C.~Cortes, N.~D. Lawrence, and K.~Q. Weinberger, Eds.\hskip 1em
  plus 0.5em minus 0.4em\relax Curran Associates, Inc., 2014, pp. 2672--2680.

\bibitem{de_rain_1}
R.~Qian, R.~T. Tan, W.~Yang, J.~Su, and J.~Liu, ``Attentive generative
  adversarial network for raindrop removal from a single image,'' \emph{CoRR},
  vol. abs/1711.10098, 2017.

\bibitem{de_blur}
O.~Kupyn, V.~Budzan, M.~Mykhailych, D.~Mishkin, and J.~Matas, ``Deblurgan:
  Blind motion deblurring using conditional adversarial networks,''
  \emph{CoRR}, vol. abs/1711.07064, 2017.

\bibitem{de_haze}
B.~Li, X.~Peng, Z.~Wang, J.~Xu, and D.~Feng, ``An all-in-one network for
  dehazing and beyond,'' \emph{CoRR}, vol. abs/1707.06543, 2017.

\bibitem{2023-Stocco-TSE}
A.~Stocco, B.~Pulfer, and P.~Tonella, ``{Mind the Gap! A Study on the
  Transferability of Virtual Versus Physical-World Testing of Autonomous
  Driving Systems},'' \emph{IEEE Transactions on Software Engineering},
  vol.~49, no.~04, pp. 1928--1940, apr 2023.

\bibitem{deeproad}
\BIBentryALTinterwordspacing
M.~Zhang, Y.~Zhang, L.~Zhang, C.~Liu, and S.~Khurshid, ``Deeproad: Gan-based
  metamorphic testing and input validation framework for autonomous driving
  systems,'' in \emph{Proceedings of ASE '18}, ser. ASE 2018.\hskip 1em plus
  0.5em minus 0.4em\relax New York, NY, USA: ACM, 2018, pp. 132--142. [Online].
  Available: \url{http://doi.acm.org/10.1145/3238147.3238187}
\BIBentrySTDinterwordspacing

\bibitem{sim2real_gans_cit}
\BIBentryALTinterwordspacing
A.~Shrivastava, T.~Pfister, O.~Tuzel, J.~Susskind, W.~Wang, and R.~Webb,
  ``Learning from simulated and unsupervised images through adversarial
  training,'' \emph{CoRR}, vol. abs/1612.07828, 2016. [Online]. Available:
  \url{http://arxiv.org/abs/1612.07828}
\BIBentrySTDinterwordspacing

\bibitem{sim2real_gans_cit_2}
\BIBentryALTinterwordspacing
J.~Hoffman, E.~Tzeng, T.~Park, J.~Zhu, P.~Isola, K.~Saenko, A.~A. Efros, and
  T.~Darrell, ``Cycada: Cycle-consistent adversarial domain adaptation,''
  \emph{CoRR}, vol. abs/1711.03213, 2017. [Online]. Available:
  \url{http://arxiv.org/abs/1711.03213}
\BIBentrySTDinterwordspacing

\bibitem{pix2pix}
P.~Isola, J.-Y. Zhu, T.~Zhou, and A.~A. Efros, ``Image-to-image translation
  with conditional adversarial networks,'' \emph{CVPR}, 2017.

\bibitem{unit}
\BIBentryALTinterwordspacing
M.~Liu, T.~M. Breuel, and J.~Kautz, ``{Unsupervised Image-to-Image Translation
  Networks},'' \emph{CoRR}, vol. abs/1703.00848, 2017. [Online]. Available:
  \url{http://arxiv.org/abs/1703.00848}
\BIBentrySTDinterwordspacing

\bibitem{disco_gan}
\BIBentryALTinterwordspacing
T.~Kim, M.~Cha, H.~Kim, J.~K. Lee, and J.~Kim, ``Learning to discover
  cross-domain relations with generative adversarial networks,'' \emph{CoRR},
  vol. abs/1703.05192, 2017. [Online]. Available:
  \url{http://arxiv.org/abs/1703.05192}
\BIBentrySTDinterwordspacing

\bibitem{cyclegan}
J.-Y. Zhu, T.~Park, P.~Isola, and A.~A. Efros, ``Unpaired image-to-image
  translation using cycle-consistent adversarial networks,'' in \emph{Computer
  Vision (ICCV), 2017 IEEE International Conference on}, 2017.

\bibitem{biagiola2023better}
M.~Biagiola, A.~Stocco, V.~Riccio, and P.~Tonella, ``Two is better than one:
  Digital siblings to improve autonomous driving testing,'' 2023.

\bibitem{2023-Stocco-EMSE}
A.~Stocco, B.~Pulfer, and P.~Tonella, ``Model vs system level testing of
  autonomous driving systems: A replication and extension study,''
  \emph{Empirical Softw. Engg.}, vol.~28, no.~3, may 2023.

\bibitem{cgan}
M.~Mirza and S.~Osindero, ``Conditional generative adversarial nets,''
  \emph{CoRR}, vol. abs/1411.1784, 2014.

\bibitem{gan_problems_1}
A.~Borji, ``Qualitative failures of image generation models and their
  application in detecting deepfakes,'' 2023.

\bibitem{inception-metric}
T.~Salimans, I.~Goodfellow, W.~Zaremba, V.~Cheung, A.~Radford, and X.~Chen,
  ``Improved techniques for training gans,'' 2016.

\bibitem{inception}
C.~Szegedy, W.~Liu, Y.~Jia, P.~Sermanet, S.~Reed, D.~Anguelov, D.~Erhan,
  V.~Vanhoucke, and A.~Rabinovich, ``Going deeper with convolutions,'' 2014.

\bibitem{fid-metric}
M.~Heusel, H.~Ramsauer, T.~Unterthiner, B.~Nessler, and S.~Hochreiter, ``Gans
  trained by a two time-scale update rule converge to a local nash
  equilibrium,'' 2018.

\bibitem{frechet}
M.~Fr{\'e}chet, ``Sur la distance de deux lois de probabilité,'' \emph{Annales
  de l'ISUP}, vol.~VI, no.~3, pp. 183--198, 1957.

\bibitem{kid-citations}
J.~Yu, X.~Xu, F.~Gao, S.~Shi, M.~Wang, D.~Tao, and Q.~Huang, ``Towards
  realistic face photo-sketch synthesis via composition-aided gans,'' 2020.

\bibitem{gram_matrix}
C.~M. Bishop, \emph{Pattern Recognition and Machine Learning}, 1st~ed.\hskip
  1em plus 0.5em minus 0.4em\relax Springer, 2006.

\bibitem{maximum-mean}
A.~Gretton, K.~M. Borgwardt, M.~J. Rasch, B.~Sch{\"o}lkopf, and A.~Smola, ``A
  kernel two-sample test,'' \emph{Journal of Machine Learning Research},
  vol.~13, pp. 723--773, 2012.

\bibitem{struct_sim}
Z.~Wang, A.~Bovik, H.~Sheikh, and E.~Simoncelli, ``Image quality assessment:
  from error visibility to structural similarity,'' \emph{IEEE Transactions on
  Image Processing}, vol.~13, no.~4, pp. 600--612, 2004.

\bibitem{cosine_sim}
F.~Zhan, Y.~Yu, K.~Cui, G.~Zhang, S.~Lu, J.~Pan, C.~Zhang, F.~Ma, X.~Xie, and
  C.~Miao, ``Unbalanced feature transport for exemplar-based image
  translation,'' 06 2021, pp. 15\,023--15\,033.

\bibitem{glcm}
S.-C. Lam, ``Texture feature extraction using gray level gradient based
  co-occurence matrices,'' in \emph{1996 IEEE International Conference on
  Systems, Man and Cybernetics. Information Intelligence and Systems (Cat.
  No.96CH35929)}, vol.~1, 1996, pp. 267--271 vol.1.

\bibitem{wd_loss}
M.~Arjovsky, S.~Chintala, and L.~Bottou, ``{W}asserstein generative adversarial
  networks,'' in \emph{Proceedings of the 34th International Conference on
  Machine Learning}, ser. Proceedings of Machine Learning Research, D.~Precup
  and Y.~W. Teh, Eds., vol.~70.\hskip 1em plus 0.5em minus 0.4em\relax PMLR,
  06--11 Aug 2017, pp. 214--223.

\bibitem{earthmover}
Y.~Rubner, C.~Tomasi, and L.~J. Guibas, ``The earth mover's distance as a
  metric for image retrieval,'' \emph{Int. J. Comput. Vision}, vol.~40, no.~2,
  p. 99–121, nov 2000.

\bibitem{kl_div}
P.~Diaconis and S.~L. Zabell, ``Updating subjective probability,''
  \emph{Journal of the American Statistical Association}, vol.~77, no. 380, pp.
  822--830, 1982.

\bibitem{hist_inter}
S.~M. Lee, J.~H. Xin, and S.~Westland, ``Evaluation of image similarity by
  histogram intersection,'' \emph{Color Research \& Application}, vol.~30,
  no.~4, pp. 265--274, 2005.

\bibitem{resnet}
K.~He, X.~Zhang, S.~Ren, and J.~Sun, ``{Deep Residual Learning for Image
  Recognition},'' 2015.

\bibitem{VGGNet}
K.~Simonyan and A.~Zisserman, ``Very deep convolutional networks for
  large-scale image recognition.''

\bibitem{perc_loss}
J.~Johnson, A.~Alahi, and L.~Fei{-}Fei, ``Perceptual losses for real-time style
  transfer and super-resolution,'' \emph{CoRR}, vol. abs/1603.08155, 2016.

\bibitem{fcn_8}
J.~Long, E.~Shelhamer, and T.~Darrell, ``Fully convolutional networks for
  semantic segmentation,'' 06 2015, pp. 3431--3440.

\bibitem{precrashreport}
N.~H. T. S.~A. U.S. Department~of Transportation, ``Pre-crash scenario typology
  for crash avoidance research,'' 2007.

\bibitem{kitti}
A.~Geiger, P.~Lenz, C.~Stiller, and R.~Urtasun, ``Vision meets robotics: The
  kitti dataset,'' \emph{International Journal of Robotics Research (IJRR)},
  2013.

\bibitem{vkitti}
A.~Gaidon, Q.~Wang, Y.~Cabon, and E.~Vig, ``Virtual worlds as proxy for
  multi-object tracking analysis,'' in \emph{CVPR}, 2016.

\bibitem{kitti_use_1}
\BIBentryALTinterwordspacing
B.~Wu, A.~Wan, X.~Yue, and K.~Keutzer, ``Squeezeseg: Convolutional neural nets
  with recurrent {CRF} for real-time road-object segmentation from 3d lidar
  point cloud,'' \emph{CoRR}, vol. abs/1710.07368, 2017. [Online]. Available:
  \url{http://arxiv.org/abs/1710.07368}
\BIBentrySTDinterwordspacing

\bibitem{kitti_use_2}
Z.~Li, X.~Wang, X.~Liu, and J.~Jiang, ``Binsformer: Revisiting adaptive bins
  for monocular depth estimation,'' 2022.

\bibitem{yolo}
\BIBentryALTinterwordspacing
J.~Redmon, S.~K. Divvala, R.~B. Girshick, and A.~Farhadi, ``You only look once:
  Unified, real-time object detection,'' \emph{CoRR}, vol. abs/1506.02640,
  2015. [Online]. Available: \url{http://arxiv.org/abs/1506.02640}
\BIBentrySTDinterwordspacing

\bibitem{unity}
``Unity3d.'' \url{https://unity.com}, 2021.

\bibitem{donkeycar}
``{Donkey Car},'' \url{https://www.donkeycar.com/}, 2021.

\bibitem{yolov3}
J.~Redmon and A.~Farhadi, ``Yolov3: An incremental improvement,'' \emph{arXiv},
  2018.

\bibitem{yolo_use_1}
\BIBentryALTinterwordspacing
D.~Wu, M.~Liao, W.~Zhang, and X.~Wang, ``{YOLOP:} you only look once for
  panoptic driving perception,'' \emph{CoRR}, vol. abs/2108.11250, 2021.
  [Online]. Available: \url{https://arxiv.org/abs/2108.11250}
\BIBentrySTDinterwordspacing

\bibitem{yolo_use_2}
A.~Sarda, S.~Dixit, and A.~Bhan, ``Object detection for autonomous driving
  using yolo algorithm,'' in \emph{2021 2nd International Conference on
  Intelligent Engineering and Management (ICIEM)}, 2021, pp. 447--451.

\bibitem{yolo_weights_web_1}
P.~Ruan, ``object-detection---yolov3,''
  \url{https://github.com/patrick013/Object-Detection---Yolov3/tree/master},
  2020.

\bibitem{yolo_weights_web_2}
\BIBentryALTinterwordspacing
openvino A.I., ``Yolo-v3-tf¶,'' 2020. [Online]. Available:
  \url{https://docs.openvino.ai/2023.1/omz_models_model_yolo_v3_tf.html}
\BIBentrySTDinterwordspacing

\bibitem{yolo_v3_speed}
U.~Nepal and H.~Eslamiat, ``Comparing yolov3, yolov4 and yolov5 for autonomous
  landing spot detection in faulty uavs,'' \emph{Sensors}, vol.~22, no.~2,
  2022.

\bibitem{coco}
T.~Lin, M.~Maire, S.~J. Belongie, L.~D. Bourdev, R.~B. Girshick, J.~Hays,
  P.~Perona, D.~Ramanan, P.~Doll{\'{a}}r, and C.~L. Zitnick, ``Microsoft
  {COCO:} common objects in context,'' \emph{CoRR}, vol. abs/1405.0312, 2014.

\bibitem{nvidia-dave2}
\BIBentryALTinterwordspacing
M.~Bojarski, D.~Del~Testa, D.~Dworakowski, B.~Firner, B.~Flepp, P.~Goyal, L.~D.
  Jackel, M.~Monfort, U.~Muller, J.~Zhang, X.~Zhang, J.~Zhao, and K.~Zieba,
  ``End to end learning for self-driving cars.'' \emph{CoRR}, vol.
  abs/1604.07316, 2016. [Online]. Available:
  \url{http://arxiv.org/abs/1604.07316}
\BIBentrySTDinterwordspacing

\bibitem{deeptest}
\BIBentryALTinterwordspacing
Y.~Tian, K.~Pei, S.~Jana, and B.~Ray, ``Deeptest: Automated testing of
  deep-neural-network-driven autonomous cars,'' in \emph{Proceedings of ICSE
  '18}, ser. ICSE '18.\hskip 1em plus 0.5em minus 0.4em\relax New York, NY,
  USA: ACM, 2018, pp. 303--314. [Online]. Available:
  \url{http://doi.acm.org/10.1145/3180155.3180220}
\BIBentrySTDinterwordspacing

\bibitem{2020-Humbatova-ICSE}
N.~Humbatova, G.~Jahangirova, G.~Bavota, V.~Riccio, A.~Stocco, and P.~Tonella,
  ``{Taxonomy of Real Faults in Deep Learning Systems},'' ser. ICSE'20.\hskip
  1em plus 0.5em minus 0.4em\relax New York, NY, USA: ACM, 2020.

\bibitem{nvidia_cnn_vision}
M.~Bojarski, P.~Yeres, A.~Choromanska, K.~Choromanski, B.~Firner, L.~D. Jackel,
  and U.~Muller, ``Explaining how a deep neural network trained with end-to-end
  learning steers a car,'' \emph{CoRR}, vol. abs/1704.07911, 2017.

\bibitem{2021-Jahangirova-ICST}
G.~Jahangirova, A.~Stocco, and P.~Tonella, ``Quality metrics and oracles for
  autonomous vehicles testing,'' in \emph{Proceedings of 14th IEEE
  International Conference on Software Testing, Verification and Validation},
  ser. ICST '21.\hskip 1em plus 0.5em minus 0.4em\relax IEEE, 2021.

\bibitem{iou}
T.~Dean, M.~A. Ruzon, M.~Segal, J.~Shlens, S.~Vijayanarasimhan, and J.~Yagnik,
  ``Fast, accurate detection of 100,000 object classes on a single machine,''
  in \emph{2013 IEEE Conference on Computer Vision and Pattern Recognition},
  2013, pp. 1814--1821.

\bibitem{Weiss2021FailSafe}
M.~Weiss and P.~Tonella, ``Fail-safe execution of deep learning based systems
  through uncertainty monitoring,'' in \emph{IEEE 14th International Conference
  on Software Testing, Validation and Verification}, ser. ICST '21.\hskip 1em
  plus 0.5em minus 0.4em\relax IEEE, 2021.

\bibitem{2024-Grewal-ICST}
R.~Grewal, P.~Tonella, and A.~Stocco, ``{Predicting Safety Misbehaviours in
  Autonomous Driving Systems using Uncertainty Quantification},'' in
  \emph{Proceedings of 17th IEEE International Conference on Software Testing,
  Verification and Validation}, ser. ICST '24.\hskip 1em plus 0.5em minus
  0.4em\relax IEEE, 2024, p. 12 pages.

\bibitem{10.5555/3045390.3045502}
Y.~Gal and Z.~Ghahramani, ``Dropout as a bayesian approximation: Representing
  model uncertainty in deep learning,'' in \emph{Proceedings of the 33rd
  International Conference on International Conference on Machine Learning -
  Volume 48}, ser. ICML '16.\hskip 1em plus 0.5em minus 0.4em\relax JMLR.org,
  2016.

\bibitem{2022-Stocco-ASE}
A.~Stocco, P.~J. Nunes, M.~d'Amorim, and P.~Tonella, ``{ThirdEye: Attention
  Maps for Safe Autonomous Driving Systems},'' in \emph{Proceedings of 37th
  IEEE/ACM International Conference on Automated Software Engineering}, ser.
  ASE '22.\hskip 1em plus 0.5em minus 0.4em\relax IEEE/ACM, 2022.

\bibitem{DBLP:journals/corr/SelvarajuDVCPB16}
R.~R. Selvaraju, A.~Das, R.~Vedantam, M.~Cogswell, D.~Parikh, and D.~Batra,
  ``Grad-cam: Why did you say that? visual explanations from deep networks via
  gradient-based localization,'' \emph{CoRR}, vol. abs/1610.02391, 2016.

\bibitem{google_tpu}
N.~P. Jouppi, G.~Kurian, S.~Li, P.~Ma, R.~Nagarajan, L.~Nai, N.~Patil,
  S.~Subramanian, A.~Swing, B.~Towles, C.~Young, X.~Zhou, Z.~Zhou, and
  D.~Patterson, ``Tpu v4: An optically reconfigurable supercomputer for machine
  learning with hardware support for embeddings,'' 2023.

\bibitem{kingma2014adam}
D.~P. Kingma and J.~Ba, ``Adam: A method for stochastic optimization,''
  \emph{arXiv preprint arXiv:1412.6980}, 2014.

\bibitem{pearson}
\BIBentryALTinterwordspacing
W.~Kirch, Ed., \emph{Pearson's Correlation Coefficient}.\hskip 1em plus 0.5em
  minus 0.4em\relax Dordrecht: Springer Netherlands, 2008, pp. 1090--1091.
  [Online]. Available: \url{https://doi.org/10.1007/978-1-4020-5614-7_2569}
\BIBentrySTDinterwordspacing

\bibitem{inception_v3}
\BIBentryALTinterwordspacing
C.~Szegedy, V.~Vanhoucke, S.~Ioffe, J.~Shlens, and Z.~Wojna, ``Rethinking the
  inception architecture for computer vision,'' \emph{CoRR}, vol.
  abs/1512.00567, 2015. [Online]. Available:
  \url{http://arxiv.org/abs/1512.00567}
\BIBentrySTDinterwordspacing

\bibitem{imagenet}
J.~Deng, W.~Dong, R.~Socher, L.-J. Li, K.~Li, and L.~Fei-Fei, ``{ImageNet: A
  Large-Scale Hierarchical Image Database},'' in \emph{CVPR09}, 2009.

\bibitem{SegFormer}
E.~Xie, W.~Wang, Z.~Yu, A.~Anandkumar, J.~M. {\'{A}}lvarez, and P.~Luo,
  ``Segformer: Simple and efficient design for semantic segmentation with
  transformers,'' \emph{CoRR}, vol. abs/2105.15203, 2021.

\bibitem{cityscapes}
M.~Cordts, M.~Omran, S.~Ramos, T.~Rehfeld, M.~Enzweiler, R.~Benenson,
  U.~Franke, S.~Roth, and B.~Schiele, ``The cityscapes dataset for semantic
  urban scene understanding,'' in \emph{Proc. of the IEEE Conference on
  Computer Vision and Pattern Recognition (CVPR)}, 2016.

\bibitem{t-test}
Student, ``The probable error of a mean,'' \emph{Biometrika}, vol.~6, no.~1,
  pp. 1--25, 1908.

\bibitem{cohen1988statistical}
J.~Cohen, \emph{Statistical power analysis for the behavioral sciences}.\hskip
  1em plus 0.5em minus 0.4em\relax Hillsdale, N.J: L. Erlbaum Associates, 1988.

\bibitem{mcpower}
Burch, Najm, Yang, and Trick, ``Mcpower: a monte carlo approach to power
  estimation,'' in \emph{1992 IEEE/ACM International Conference on
  Computer-Aided Design}, 1992, pp. 90--97.

\bibitem{2020-Haq-ICST}
F.~U. Haq, D.~Shin, S.~Nejati, and L.~Briand, ``Comparing offline and online
  testing of deep neural networks: An autonomous car case study,'' in
  \emph{Proceedings of 13th IEEE International Conference on Software Testing,
  Verification and Validation}, ser. ICST '20.\hskip 1em plus 0.5em minus
  0.4em\relax IEEE, 2020.

\bibitem{tool}
``{Replication package},''
  \url{https://github.com/ast-fortiss-tum/I2I-quality-metrics-study.git}, 2023.

\bibitem{10.1145/3368089.3409743}
S.~Garc\'{\i}a, D.~Str\"{u}ber, D.~Brugali, T.~Berger, and P.~Pelliccione,
  ``{Robotics Software Engineering: A Perspective from the Service Robotics
  Domain},'' in \emph{Proceedings of the 28th ACM Joint Meeting on European
  Software Engineering Conference and Symposium on the Foundations of Software
  Engineering}, ser. ESEC/FSE 2020.\hskip 1em plus 0.5em minus 0.4em\relax USA:
  Association for Computing Machinery, 2020, p. 593–604.

\bibitem{hu2023sim2real}
X.~Hu, S.~Li, T.~Huang, B.~Tang, and L.~Chen, ``Sim2real and digital twins in
  autonomous driving: A survey,'' \emph{arXiv preprint arXiv:2305.01263}, 2023.

\bibitem{schwarz2022role}
C.~Schwarz and Z.~Wang, ``The role of digital twins in connected and automated
  vehicles,'' \emph{IEEE Intelligent Transportation Systems Magazine}, vol.~14,
  no.~6, pp. 41--51, 2022.

\bibitem{kalapos2020sim}
A.~Kalapos, C.~G{\'o}r, R.~Moni, and I.~Harmati, ``Sim-to-real reinforcement
  learning applied to end-to-end vehicle control,'' in \emph{2020 23rd
  International Symposium on Measurement and Control in Robotics
  (ISMCR)}.\hskip 1em plus 0.5em minus 0.4em\relax IEEE, 2020, pp. 1--6.

\bibitem{deepbillboard}
H.~Zhou, W.~Li, Y.~Zhu, Y.~Zhang, B.~Yu, L.~Zhang, and C.~Liu, ``Deepbillboard:
  Systematic physical-world testing of autonomous driving systems,'' 2018.

\bibitem{domain_rando}
X.~Chen, J.~Hu, C.~Jin, L.~Li, and L.~Wang, ``Understanding domain
  randomization for sim-to-real transfer,'' \emph{CoRR}, vol. abs/2110.03239,
  2021.

\bibitem{fathy2006review}
H.~K. Fathy, Z.~S. Filipi, J.~Hagena, and J.~L. Stein, ``Review of
  hardware-in-the-loop simulation and its prospects in the automotive area,''
  in \emph{Modeling and simulation for military applications}, vol. 6228.\hskip
  1em plus 0.5em minus 0.4em\relax SPIE, 2006, pp. 117--136.

\bibitem{DBLP:journals/corr/abs-2107-07364}
D.~Parthasarathy and A.~Johansson, ``{SilGAN: Generating driving maneuvers for
  scenario-based software-in-the-loop testing},'' \emph{CoRR}, vol.
  abs/2107.07364, 2021.

\bibitem{physgan}
Z.~Kong and C.~Liu, ``{Generating Adversarial Fragments with Adversarial
  Networks for Physical-world Implementation},'' \emph{CoRR}, vol.
  abs/1907.04449, 2019.

\bibitem{DBLP:journals/corr/abs-2005-03844}
Z.~Yang, Y.~Chai, D.~Anguelov, Y.~Zhou, P.~Sun, D.~Erhan, S.~Rafferty, and
  H.~Kretzschmar, ``{SurfelGAN: Synthesizing Realistic Sensor Data for
  Autonomous Driving},'' \emph{CoRR}, vol. abs/2005.03844, 2020.

\bibitem{empirical_2}
S.~Arora and Y.~Zhang, ``Do gans actually learn the distribution? an empirical
  study,'' \emph{CoRR}, vol. abs/1706.08224, 2017.

\bibitem{huang2018an}
G.~Huang, Y.~Yuan, Q.~Xu, C.~Guo, Y.~Sun, F.~Wu, and K.~Weinberger, ``An
  empirical study on evaluation metrics of generative adversarial networks,''
  2018.

\bibitem{2020-Stocco-ICSE}
A.~Stocco, M.~Weiss, M.~Calzana, and P.~Tonella, ``Misbehaviour prediction for
  autonomous driving systems,'' in \emph{Proceedings of 42nd International
  Conference on Software Engineering}, ser. ICSE '20.\hskip 1em plus 0.5em
  minus 0.4em\relax ACM, 2020, p. 12 pages.

\bibitem{2020-Stocco-GAUSS}
A.~Stocco and P.~Tonella, ``Towards anomaly detectors that learn
  continuously,'' in \emph{Proceedings of 31st International Symposium on
  Software Reliability Engineering Workshops}, ser. ISSREW 2020.\hskip 1em plus
  0.5em minus 0.4em\relax IEEE, 2020.

\bibitem{2021-Stocco-JSEP}
------, ``Confidence-driven weighted retraining for predicting safety-critical
  failures in autonomous driving systems,'' \emph{Journal of Software:
  Evolution and Process}, 2021.

\end{thebibliography}

\end{document}